\documentclass{article}

\usepackage{cite}
\usepackage[namelimits]{amsmath} 
\usepackage{amssymb}             
\usepackage{geometry}

\usepackage{graphicx}
\usepackage[lined,linesnumbered,ruled,vlined,commentsnumbered]{algorithm2e}
\usepackage{subfig}
\usepackage[title]{appendix}
\usepackage{CJK}
\usepackage{color}
\usepackage[misc]{ifsym}
\usepackage[colorlinks=true,linkcolor=blue,anchorcolor=blue,citecolor=blue,urlcolor=blue]{hyperref}
\usepackage[marginal]{footmisc}
\usepackage[T1]{fontenc}
\usepackage[utf8]{inputenc}
\usepackage{authblk}

\newcommand{\subsubsubsection}[1]{\paragraph{#1}\mbox{}\\} 

\newtheorem{definition}{Definition}
\newtheorem{theorem}{Theorem}
\newtheorem{proof}{Proof}
\newtheorem{remark}{Remark}
\newtheorem{problem}{Problem}
\newtheorem{example}{Example}
\providecommand{\keywords}[1]{\small\textbf{\textit{Keywords---}} #1}
\begin{document}

\title{Similarity-driven and Task-driven Models for Diversity of Opinion in Crowdsourcing Markets
}

\author[b]{Chen Jason Zhang}
\author[a]{Yunrui Liu}
\author[a]{Pengcheng Zeng \thanks{Corresponding author: zengpch@shanghaitech.edu.cn}}
\author[b]{Ting Wu}
\author[c]{Lei Chen}
\author[c]{Pan Hui}
\author[b]{Fei Hao}
\affil[a]{ShanghaiTech University, Shanghai, China}
\affil[b]{The Hong Kong Polytechnic University, Hong Kong SAR, China}
\affil[c]{The Hong Kong University of Science and Technology, Hong Kong SAR, China}
\date{}
\maketitle
\footnote{$\quad$ Chen Jason Zhang, Yunrui Liu and Pengcheng Zeng contribute equally to this work.}
\begin{abstract}
The recent boom in crowdsourcing has opened up a new avenue for utilizing human intelligence in the realm of data analysis. This innovative approach provides a powerful means for connecting online workers to tasks that cannot effectively be done solely by machines or conducted by professional experts due to cost constraints. Within the field of social science, four elements are required to construct a sound crowd - Diversity of Opinion, Independence, Decentralization and Aggregation. However, while the other three components have already been investigated and implemented in existing crowdsourcing platforms, `Diversity of Opinion' has not been functionally enabled yet.\\
From a computational point of view, constructing a wise crowd necessitates quantitatively modeling and taking diversity into account. There are usually two paradigms in a crowdsourcing marketplace for worker selection: building a crowd to wait for tasks to come and selecting workers for a given task. We propose similarity-driven and task-driven models for both paradigms. Also, we develop efficient and effective algorithms for recruiting a limited number of workers with optimal diversity in both models. To validate our solutions, we conduct extensive experiments using both synthetic datasets and real data sets.
\end{abstract}

\section{Introduction}
\label{background}
Recently, with the emergence of crowdsourcing platforms, such as Amazon Mechanical Turk \cite{Ref3} and CrowdFlower \cite{Ref4}, more and more applications are utilizing human intelligence in processing various tasks that are either too difficult to be solved only by computers alone or too expensive to employ experts to perform. For example, data gathering can be done implicitly, through crowdsourced sensing and on-line behaviour collection, or explicitly, by sending targeted information requests to the crowd. Given another example from an analytical perspective, human input can be used to address computationally difficult tasks such as entity resolution \cite{Ref34}, schema matching \cite{Ref35} and the like. In recent years, the higher-efficiency crowdsourcing models like CDB \cite{Ref40} and SPR \cite{Ref44} were proposed. They were able to further expand the scope and frequency of use of crowdsourcing.\\
\indent Nowadays, the emergence and development of large language models (LLMs), which can be considered as a manifestation of collective wisdom stemming from the contributions of numerous individuals, have significantly influenced crowdsourcing. However, LLMs have a dual impact on crowdsourcing. On one hand, LLMs can offer innovative solutions for crowdsourcing tasks by providing ideas and methodologies. On the other hand, they may also simulate responses akin to those of ordinary individuals, potentially leading to confusion within crowdsourcing platforms \cite{Ref54}.\\
\indent Furthermore, the advent of LLMs does not signify the obsolescence of crowdsourcing. Crowdsourcing, fundamentally rooted in human intellect and capability, persists due to the diverse nature of its demands. Given the wide array of tasks encompassed within crowdsourcing, specific assignments necessitate individuals possessing particular attributes, skills, and preferences. In certain scenarios, the investment required to train an LLM for a specific task may exceed the effort of sourcing a suitably qualified crowd for crowdsourcing. Simultaneously, crowdsourcing offers invaluable benefits to LLMs. It serves as a source of data, facilitates feedback on generated outcomes, and enables supervision. Additionally, crowdsourcing plays a pivotal role in LLM development, as the training data for such models are often constructed artificially. Therefore, crowdsourcing can help the advancement of large language models.
\\
\indent Though humankind is intelligent, meanwhile, they are also erroneous and greedy, which makes the quality of crowdsourcing results quite questionable. Therefore, it is important to select the `right' workers to build a wise crowd to guarantee the quality \cite{Ref55}. Then one crucial question to address is `What are the elements of a wise crowd?'. Fortunately, this question has been thoroughly studied in the field of social science and many detailed answers have been given. One of the most recognized answers, from \cite{Ref31} with over 5,000 citations, points out that four elements are essential to form a wise crowd, which are:
\begin{itemize}
    \item 1. Diversity of Opinion - Each person should have private information even if it's just an eccentric interpretation of the known facts.
    \item 2. Independence - People's opinions aren't determined by the opinions of those around them.
    \item 3. Decentralization - People are able to specialize and draw on local knowledge.
    \item 4. Aggregation - Some mechanism exists for turning private judgements into a collective decision.
\end{itemize}\par
\indent Therefore, in order to construct a wise crowd, we need to make sure that the constructed crowd satisfies the above four elements. From the perspective of crowdsourcing systems, \textit{independence} and \textit{decentralization} are easy to achieve, by providing a free and independent channel for each individual worker, that is, a means to enable each worker to answer questions based on personal specialism without being aware of other workers. Existing crowdsourcing platforms, such as AMT and CrowdFlower, work precisely in this way. Concerning \textit{aggregation}, various mechanisms have been proposed already, such as majority voting \cite{Ref10}, to achieve a target overall reliability. However, to the best of our knowledge, how to ensure the \textit{diversity of opinion} in constructing a wise crowd has not been studied from algorithmic perspectives before. Thus, in this paper, we address the algorithmic optimizations towards the \textit{diversity of opinion} for crowd construction.
\subsection{When Diversity Trumps Ability}
\label{Diversity trumps ability}
The effect of diversity differs depending on the corresponding crowdsourced tasks, as pointed out in \cite{Ref23}. In particular, for \textit{problem-solving tasks}, diversity is the essential factor affecting the performance of a crowd, and it is even much more important than the average ability of individuals. This phenomenon was discovered and verified in \cite{Ref24}, and referred to the `Diversity Trumps Ability Theorem', which makes the observation that diverse groups of problem solvers - groups of people with diverse tools consistently outperformed groups of the best and the brightest. People with high abilities are often trained in the same institutions, tend to possess similar perspectives and apply similar problem-solving techniques, or heuristics. Many problems do not succumb to a single heuristic, or even a set of similar ones. This is why a diverse crowd functions better than a few experts. Intuitively, if two groups are formed, one random (and therefore diverse) and one consisting of the best individual performers, the first group almost always did better.\\
\indent The diversity of opinion holds significant practical importance for several reasons. A diverse spectrum of opinions can challenge conventional modes of thinking, fostering the generation of more innovative solutions. For instance, within the business realm, teams comprising individuals from diverse backgrounds often exhibit greater capacity for innovation compared to homogeneous groups. This phenomenon arises from the varied experiences, skills, and perspectives that team members bring to the collaborative process, facilitating a multi-faceted approach to problem-solving and the development of more comprehensive strategies. Furthermore, the diversity of opinion plays a crucial role in the decision-making process. It ensures that decisions undergo thorough scrutiny from multiple vantage points, mitigating the risk of succumbing to `groupthink', where the pursuit of consensus supersedes objective evaluation of alternative options. As a result, decision-making processes are fortified, yielding more robust and well-informed decisions that account for a broader spectrum of potential impacts and outcomes.
\\
\indent This theorem ends up indirectly providing convincing arguments as to why - under certain conditions - citizens may outperform elected officials and experts \cite{Ref23}.
\begin{table}[ht]
\centering
\caption{MEANINGS OF SYMBOLS USED}
\begin{tabular}{ll}
\hline
Notation & Description\\
$w(w_i)$ & a crowdsourcing worker\\
$Sim(w_i,w_j)$ &  the pairwise similarity between $w_i$ and $w_j$\\
$Div(C)$ & the diversity of a crowd C of workers\\
$\theta_1(\theta_0)$ & the number of positive (negative) workers\\ & to be enlisted with positive (negative)\\
$t_i$ & the opinion of worker $w_i$\\
$Pr(t = 1$ or $0)$ & the probability of $t$ satisfying or \\& dissatisfying $P$\\
$N$ & the set of candidate workers to be selected\\
$k$ & the number of workers to be enlisted\\
$S$ & the set of workers to be selected, $|S| = k$\\
$\theta_2$ & $\theta_2 = k - \theta_0$\\
$\tau(S)$ & the probability of at least $\theta_1 (\theta_0)$ workers\\ & existing in S\\
$T_0$ & $T_0 = \sum_{t\in S}t$, following Poisson Binomial \\ & distribution
\end{tabular}
\end{table}
\subsection{Two Basic Models for Diversity of Opinion}
\label{Models' intro}
From a computational perspective, in order to build a wise crowd, we are interested in quantitatively modeling the diversity, and take it into consideration for constructing a crowd. In a crowdsourcing marketplace, we usually encounter two basic paradigms for worker selection: building a crowd that will wait for tasks to come or selecting workers for a given task. We propose models for both of the
paradigms.
\subsubsection{Similarity-driven Model (S-model)}
\label{S-model intro}
When there is no explicit query, we resort to the pairwise similarity of workers to model the diversity of opinion. In particular, we model the similarity of a pair of workers as a similarity score value (high value indicates high similarity), and use the negative value of average pairwise similarity to quantify the overall diversity. Intuitively, the lower the average similarity, the higher the diversity.\\
\indent S-model can be applied to crowdsourcing scenarios which do not have explicit queries when constructing a crowd and require quick responses when a query arrives. For example, diners may comment on a restaurant through Foursquare \cite{Ref1}, whereas iPhone users may post ratings of the applications that they have downloaded from the Apple Store. Such data is highly valuable for product creators (usually a company) : as ratings and reviews have a significant impact on sales; and companies can analyze ratings and review trends to adjust overall marketing strategies, improve customer service, and fine-tune merchandising and so on. However, in current web-based commenting systems, product creators must passively wait for reviewers to visit the commenting systems to provide their comments and ratings. Hence, product creators may have to wait a long time to receive a satisfactory number of reviews. These drawbacks with existing commenting systems motivate the quest for effective methods to actively invite a group of reviewers prior to the arrival of the query.
\subsubsection{Task-driven Model (T-model)}
\label{T-model intro}
Another common scenario is that a requester has a specific query, and enlists workers to join the crowd to answer it. In such a paradigm, we are able to analyze the diversity of workers according to the content of the query. Regarding the given query, we model the opinion of each worker as a probability ranging from 0 to 1, which indicates opinions from negative to positive, respectively. To guarantee the desirable diversity of opinion, we allow a user to set up the demand on the number of workers with positive (negative) opinions. Therefore, the optimization issue is to maximize the probability that the user’s demand is satisfied.\\
\indent T-model captures essence of diversity for a wide class of crowdsourcing scenarios. A typical example application, which is initiated and currently operated by the US government \cite{Ref2}, is an \textit{online petitioning system} enabling participants to propose, discuss and sign political petitions. To determine whether a petition is significant enough to get a response from the White House, the current mechanism is simply a threshold of the number of signatures (currently 100,000), indicating the number of people who support the petition. However, to analyze a particular petition fairly, it would be more constructive if opinions from both the proposition and the opposition are taken into consideration. So guided by the T-model, the government may actively collect online comments on both sides of the petition, which is more constructive for further governmental processing.
\subsection{Challenges and Contributions}
\label{challenges and the text part's intro}
As \textit{diversity} is a loosely defined concept, the first main challenge is quantitatively measuring the diversity among candidate workers. Another main challenge to be addressed is to design effective and efficient algorithms for worker selection with the consideration of the diversity of opinions. To address these two challenges, we propose effective measures to estimate the diversity of the crowd under two common scenarios, S-model and T-model, respectively, and propose effective approximation algorithms for crowd selection. To summarize, this paper has made the following contributions:\\
\begin{itemize}
\item 1. In Section \ref{S-model part}, we study the crowd selection problem under S-model, and propose an efficient (1 + $\epsilon$) approximation algorithm for finding a crowd with the highest diversity.
\item 2. In Section \ref{T-model part}, we study the crowd selection problem under the T-model, prove its NP-hardness, and provide a solution based on distribution approximations or an exact method.
\item 3. In section \ref{Expermental evaluation and Case study}, we present our experimental evaluation of the performances of T-model and S-model, and conduct a case study to exhibit the goodness of
crowds selected by our proposed models.
\item 4. In Sections \ref{Related Work} and \ref{Conclusion and Future work}, we discuss related works and conclude the paper.
\end{itemize}
\indent Compared with the conference version, this paper has following new contributions:
\begin{itemize}
\item 1. In this paper, we proposed two methods, MIN-SIM (i.e. initially selecting workers with the smallest $Sim$ value) and MIN-SUM (i.e. initially selecting workers with the smallest sum of related $Sim$ values) for evaluating the effectiveness of the algorithm for S-model and added one more synthetic data and one more real data for testing these two methods (see Sections \ref{Approximation the S-model} and \ref{Experiments on S-model}). These are very much needed to supplement the deficiencies in the discussion of the S-model in previous conference paper.
\item 2. In our previous conference paper, we proposed two methods to optimize the process of worker selection under Task-driven model towards the diversity of opinion in crowdsourcing markets. These methods include Poisson approximation and Binomial approximation. Both of them are not very accurate and not fast, since they approximate the objection function twice. In this paper, we introduced two new methods that perform much better. The first one is normal approximation with the simulated annealing algorithm (see Section \ref{Normal Approximation and Simulated Annealing method}), which only approximate the objection function once, and is much faster than the previous two methods. The second one is the method DFT-CF with the simulated annealing algorithm (see Section \ref{exact DFT-CF method}), which is an exact method and is more accurate than the previous three methods.
\item 3. We conducted a series of experiments on T-model using both synthetic data (see Section \ref{Experiments on synthetic data}) and real data (see Section \ref{Experiments on real data}) to evaluate the performance of the two newly introduced methods. The results consistently demonstrate that these two new methods exhibit stable and superior performance compared to the two methods proposed in our previous paper (see Figs \ref{fig:7}-\ref{fig:9}). The detailed explanations for these phenomenons are given in Section \ref{Experiments on T-model}.
\end{itemize}
\section{SIMILARITY-DRIVEN MODEL}
\label{S-model part}
In this section, we formally introduce the model, and propose efficient algorithms to enlist workers.

\subsection{Model and Definitions}
\label{S-model's definition}
\begin{figure}
\centering
  \includegraphics[scale=0.42]{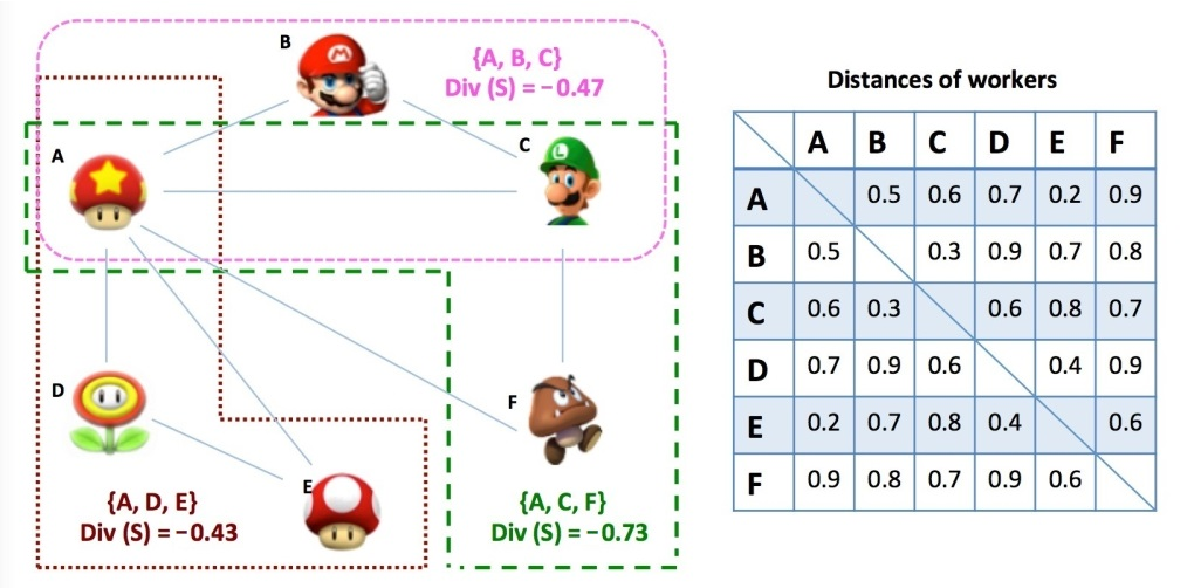}
\caption{Find 3 workers with highest diversity. The similarity matrix in the right panel demonstrates the distances among 6 workers who are shown in the left panel, and the left panel shows three concrete cases of computing $Div(S)$ for three workers based on the pairwise similarity values.}
\label{fig:1}       
\end{figure}

We first need to design a computational model to depict the crowd diversity for the worker selection problem. Under the similarity-driven model, each pair of workers is associated with a value which describes their pairwise similarity. We aim to select $k$ workers out of $n$ candidates, such that the average pairwise distance is maximized (i.e. the average similarity is minimized) \cite{Ref56}.\\
\indent We formally present the model with the following definitions.
\begin{definition}[PAIRWISE SIMILARITY]
\textit{For a given set of potential crowdsourcing workers $W$, the diversity of any two workers is computed by a pairwise similarity function $Sim(w_i, w_j)$ where $w_i, w_j \in W$.}
\end{definition}
\begin{definition}[CROWD DIVERSITY]
\textit{Given a crowd of workers $C = \{w_1,w_2,..., w_{|C|}\}$, a pairwise similarity function $Sim(.)$, the diversity of the crowd is defined as the negative value averaged pairwise similarity, that is,}
\end{definition}
\begin{equation}
    Div(C) = \frac{-\sum_{w_i,w_j\in C\wedge i\neq j}Sim(w_i,w_j)}{|C|}
\end{equation}
\begin{remark}
For the sake of generality, we consider $Sim(.)$ here as an abstract function, which measures the \textit{similarity} between two workers. In Section \ref{Similarity measurement of S-model}, we list a number of popular methods to quantify $Sim(.)$. Aside from these measurements, we can also plug in any reasonable diversity measurements. In our model, users may also design appropriate similarity functions depending on the data structure and application requirements.
\end{remark}
\indent Essentially, we are interested in finding a subset of candidate workers with the maximal diversity, using the cardinality constraint. We formally define this optimization problem as follows.
\begin{problem}[DIVERSITY MAXIMIZATION] \textit{For a given set of potential crowdsourcing workers $W$, each worker $w_i \in W$, an integer $k$, we aim to find a subset $C \subseteq W$ such that $|C| = k$ and $Div(C)$ is maximized, that is,}
\end{problem}
\begin{equation}
     \mathop{\arg\max}_{C\subseteq W,|C|=k} Div(C)\
\end{equation}
\begin{example} Fig.\ref{fig:1} illustrates an example with 6 workers and their pairwise similarity values. We aim to select three of them, to maximize the crowd diversity. To find 3 workers with highest diversity, Table 2 enumerates all the possible selections and the associated crowd diversities, and it is clear the optimal selection is $<A,D,E>$, with the highest diversity -0.433.
\end{example}
\begin{table}[ht]\centering
\begin{tabular}{|c|c|c|c|c|c|}
\hline
Crowd & $Div(S)$ & Crowd & $Div(S)$ & Crowd & $Div(S)$\\ \hline
A,B,C & -0.467 & A,B,D & -0.7 & A,B,E & -0.467 \\\hline
A,B,F & -0.733 & A,C,D & -0.633 & A,C,E & -0.533 \\\hline
A,C,F & -0.733 & \textbf{A,D,E} & \textbf{-0.433} & A,D,F & -0.833\\\hline
A,E,F & -0.567 & B,C,D & -0.6 & B,C,E & -0.6\\\hline
B,C,F & -0.6 & B,D,E & -0.667 & B,D,F & -0.867\\\hline
B,E,F & -0.7 & C,D,E & -0.6 & C,D,F & -0.73\\\hline
C,E,F & -0.7 & D,E,F & -0.633 & & \\\hline
\end{tabular}
\caption{$Div(S)$ values of all 3-worker combinations in Fig.1}
\end{table}
\subsection{Similarity Measurement of S-model}
\label{Similarity measurement of S-model}
We assume we are given a set \(\mathcal{T}\) of historical tasks and a set \(\mathcal{W}\) of workers. Each Task $t \in \mathcal{T}$ is associated with a unique identifier $t_{id}$ and a set of workers $t_W \subseteq \mathcal{W}$ who have worked on $t$. A record $e$ is a triple of the form [$t_{id},w_{id},features$] where $w_{id}$ is a unique identifier of the worker and $features$ contain certain useful information (e.g. correctness, latency, submission time, etc.) which this record refers. The set of all records belonging to a worker $w$ forms the experience of the worker denoted by \textit{experience(w)}. Without loss of generality, we assume a worker has at most one record per task.\\
\indent For each task $t$, we characterize it with a set of attributes such as category, complexity, workload, requester and nature (e.g. problem solving task, survey). Similarly, a worker $w$ could carry demographic information such as gender, age, expertises, occupation and geographic location.
\subsubsection{Pairwise Relevance}
\label{Pairwise relevance}
In a typical crowdsourcing environment, \textit{relevance} between a task and a candidate worker serves as an important criterion to guarantee the quality of the crowdsourced results. Therefore, we first introduce the definition and measurement of relevance before formally defining the concept of diversity.
\begin{definition}[PAIRWISE RELEVANCE]\textit{For a given set of potential workers \(\mathcal{W}\) and tasks \(\mathcal{T}\) , the relevance between any worker and task is computed by a given function $Rel(w_i,t_i) =  1/d_{rel}(w_i,t_i)$, where $w_i \in \mathcal{W},$ $t_i \in \mathcal{T}$.}\\\\
\end{definition}
\indent Given a task $t_i$ and a threshold radius $r$, we define the set of worker relevant to $t$ as the set of workers $w_i \in W$ within the relevant distance $r$ from $t_i$, e.g. $Rel(w_i,t_i) \leq r$. For example, the distance between a task and a worker (represented by their sets of features $x$ and $y$) could be computed by \textit{Jaccard distance}, e.g. $d_{rel}(x,y)$ = 1 - $Jaccard(x,y)$. In this paper, features are extracted from the descriptions of tasks and profile of workers by running Porter Algorithm \cite{Ref28}.
\subsubsection{Pairwise Profile-Based Diversity}
\label{Pairwise profile-based diversity}
Intuitively, we define diversity between two workers $w_i$ and $w_j$ as a function of entities extracted from their profiles.
\begin{definition}[PAIRWISE SIMILARITY] \textit{For a given set of potential crowdsourcing workers $W$, the diversity of any two workers is computed by the similarity function $Sim(w_i,w_j) = Jaccard(w_i,w_j )$, where $w_i,w_j \in W$.}\\\\
\end{definition}
\indent Thus, two workers maybe similar because they have the same gender and age, but still different(diverse), if one is living in Hong Kong and the other in New York.
\subsubsection{Pairwise Experience-Based Diversity}
\label{Pairwise experience-based diversity}
For a more sophisticated measurement, we denote the \textit{experience} $\varepsilon$ as a collection of historical records of each worker. Diversity between two workers $w_i$ and $w_j$ is defined as a function of experience engaged by works through their activities on the historic tasks. There are two steps for inferring pairwise experience-based diversity of two workers.
\subsubsubsection{Probabilistic Topic Model}
\label{Probabilistic topic model}
We use a probabilistic model to model user’s experience $E_i$ as a unordered collection of words (a.k.a. bag of words). Such collection of words (i.e. task identifier, task features, etc.) can be extracted from the records of different tasks that the worker has been performed. Specifically, we use a mixture model in which each component corresponds to one of $K$ different topics. Let $\pi_k$, for $k$ = 1,...,$K$, denote the prior probability that a collection contains topic $T_k$. For each topic, there is a corresponding multinomial distribution over the $M$ distinct words in all collections. Let $\mu_{kj}$, for $k$ = 1,...,$K$, $j$ = 1,...,$M$, denote the probability that topic $T_k$ contains word $\omega_j$ in all collections. Suppose a collection $U_i$ contains a total $N_i$ words in which each word is generated i.i.d from the mixture model above. The number of occurrences of word $\omega_j$ in $E_i$ is equal to $n_{ij}$, which follows that $\sum^M_{j=i} n_{ij} = N_i$. We assume there are $N$ i.i.d collections denoted by $E_1,E_2,...,E_N$ that associated with $N$ users.\\
\indent Let $\Phi = (\pi_k,\mu_{kj})$ denote the model parameters. We estimate $\Phi$ using EM, the E-step computes for each collection $D_i$ the posterior probability that $D_i$ belongs to topic $T_k$ given the model parameters $\Phi^t$ of the previous iteration. We can apply Bayes' rule to express $P(T_k|E_i,\Phi^t)$ as
\begin{equation}
\begin{aligned}
    p(T_k|E_i,\Phi^t) &= \frac{P(T_k)P(E_i|T_i,\Phi^t)}{\sum_{l=1}^{K}P(T_l)P(E_i|T_i,\Phi^t)}\\
    &=\frac{\pi_k^t\prod_{j=1}^M(\mu_{kj})^{n_{ij}}}{\sum_{l=1}^{K}\pi_l^t\prod_{j=1}^M(\mu_{lj}^t)^{n_{ij}}}
\end{aligned}
\end{equation}
\indent In the M-step, to maximize $\Psi(\Phi|\Phi^t)$ w.r.t $\Phi$ to obtain the next estimate $\Phi^{t+1}$, we can obtain
\begin{equation}
    \pi_k^{t+1} = \frac{1}{N}\sum_{i=1}^{N}h_K^{(i)}
\end{equation}
\indent We note that there are $K$ constraints due to the multinomial distribution for the $K$ topics:
\begin{equation}
    \sum_{j=1}^{M} \mu_{kj} = 1\quad k = 1,...,K
\end{equation}
\indent To solve a constrained optimization problem, we introduce K Lagrange
multipliers.
\begin{equation}
    \frac{\partial}{\partial\mu_{kj}}[\sum_{i=1}^{N}\sum_{k=1}^Kh_k^{(i)}logP(E_i|T_k,\Phi)-\sum_{k=1}^{K}\lambda_k(\sum_{j=1}^M\mu_{kj}-1)]=0
\end{equation}
where $h_k^{(i)}$ denotes $P(T_k|E_i,\Phi^t)$. This gives
\begin{equation}
    \mu_{kj}^{t+1}=\frac{\sum_{i=1}^Nh_{k}^{(i)}n_{ij}}{\sum_{j^{\prime}=1}^{M}\sum_{i=1}^{N}h_k^{(i)}n_{ij^{\prime}}}
\end{equation}
The EM algorithm converges to a stationary point of the likelihood function. Then we obtain the probabilistic topic distribution, which is denote by $w_i.\varphi$, of each worker.
\subsubsubsection{Worker Distance Function}
\label{worker distance function}
Given two workers $w_i,w_j \in \mathcal{W}$, the topic distance between two workers is defined as
\begin{equation}
    D(w_i,w_j) = KL(w_i.\varphi||w_j.\varphi)
\end{equation}
where $KL(.)$ measures the distance between the topic distributions $w_i.\varphi$ and $w_j.\varphi$, i.e.
\begin{equation}
    KL(w_i.\varphi||w_j.\varphi) = \sum_iPr(w_i.\varphi(i))\log\frac{Pr(w_i.\varphi(i))}{Pr(w_j.\varphi(i))}
\end{equation}
Then we have $Sim(w_i,w_j) = -D(w_i,w_j)$.
\subsubsection{Discussion on two diversities.}
Here we discuss the differences between the two diversities mentioned above and their respective advantages and disadvantages. We abbreviate pairwise profile-based diversity and experience-based diversity as PD and ED respectively. PD is a method of measuring similarity through the similarity of information between two workers. Its calculation method is to directly calculate the percentage of the number of consistent features between two workers to the total number of selected features. ED uses the EM algorithm to calculate the distribution of topics contained in workers' historical information and words in reports, and calculates the KL divergence of all topics between two workers to obtain similarity.\\
\indent The information used by the two methods is different. The calculation of PD directly uses the similarity of features and is more inclined to use personal information and classification features (such as favorite color, residence, education); the calculation of ED uses the similarity of text topics, using longer text information a little more (e.g., reports, feedback on issues, survey results with open-ended questions).\\
\indent The calculation of PD is simpler and faster, and the collection of relevant information is more convenient. However, there is no weight in its calculation. When different features reflect different degrees of similarity, this method cannot reflect this well. In addition, the similarity between feature options cannot be well handled (such as age).\\
\indent The calculation of ED is relatively complex, requires a large amount of data, and is much larger than PD in terms of time and space. However, changes in language habits and workers' attitudes, tendencies, and emotions can be reflected in ED. Such analysis can better take into account workers' personalities and resumes.
\subsection{NP-hardness}
\label{NP problem under S-mdoel}
Unfortunately, the diversity maximization problem under S-model is NP-hard, as stated in the following theorem.
\begin{theorem}
\textit{The diversity maximization problem is NP-hard.}
\end{theorem} 
\begin{proof}
First, we reduce the diversity maximization problem to a subset version: relaxing the constant from $|S| = k$ to be $|S| \leq k$. The reduction is correct because, if a polynomial algorithm A solves the crowd selection problem, then we can solve this by calling A \textit{k} times, setting $|S| = 1,2,...,k$.\\
\indent Next, we construct a special case of the diversity maximization problem, namely the crowd selection problem. We reach the NP-hardness of crowd selection problem by proving the crowd selection problem is NP-hard. With a trivial reduction, the crowd selection problem becomes an nth-order Knapsack Problem according to Formula 21. Following the proof by H. Kellerer, et al in \cite{Ref19}, we prove the hardness of nOKP. \\
\indent An nth-order Knapsack Problem(nOKP) is a Knapsack problem whose objective function has the form as follows:
\begin{equation}
    optimize \sum_{i_1\in n}\sum_{i_2\in n}\cdot\cdot\cdot\sum_{i_n\in n}V[i_1,i_2,\cdot\cdot\cdot,i_n]\cdot x_1x_2\cdot\cdot\cdot x_n
\end{equation}
where $V[i_1,i_2,\cdot\cdot\cdot,i_n]$ is an n-dimensional vector indicating the profit achieved if objects $[i_1,i_2,\cdot\cdot\cdot,i_n]$  are concurrently selected. Given an instance of a traditional KP, we can construct an nOKP instance by defining the profit n-dimensional vector as $V[i,i,\cdot\cdot\cdot,i] = p_i$ and $V[otherwise] = 0$ for all $i$, where $p_i$ is the profit in a traditional KP. The weight vector and objective value remain the same. \hfill $\square$
\end{proof}
\subsection{Approximation Algorithm}
\label{Approximation the S-model}
In the previous section, we show that the diversity maximization problem is NP-hard. Therefore, we are interested in developing fast approximation algorithms.\\
\indent Now we revisit the optimization function defined in Definition 2:
$Div(C) = \frac{-\sum_{w_i.w_j\in C\wedge i\neq j}Sim(w_i,w_j)}{|C|}$, in which $|C|$ is a fixed value, indicating the number of workers to be selected. Hence, the goal is actually to maximize $-\sum_{w_i.w_j\in C\wedge i\neq j}Sim(w_i,w_j)$, which we use $Sum(C)$ to denote. As a result, we have
\begin{equation}
    Sum(C) = -\sum_{w_i.w_j\in C\wedge i\neq j}Sim(w_i,w_j)
\end{equation}
Then, the optimization is equivalently transformed as
\begin{equation}
    \mathop{\arg\max}_{C\subseteq W,|C|=k} Sum(C)\
\end{equation}
\indent Furthermore, we discover that the optimization function $Sum(.)$ is a submodular function of the set of candidate workers $W$.\\
\indent A function $f$ \emph{is} submodular if
\begin{equation}
    f(A\cup \{a_1\}) + f(A\cup \{a_2\}) \geq f(A\cup \{a_1, a_2\}) + f(A)
\end{equation}
for any $A$ and $a_1, a_2 \notin A$. Submodularity implies the property of diminishing marginal returns. Intuitively, in our problem, this says that adding a new worker would lead to an enhanced improvement if there were less workers already in the crowd. The problem of selecting a k-element subset maximizing a sub-modular function can be approximated with a performance guarantee of $(1 - 1/e)$, by iteratively selecting the best element given the ones selected so far.\\
\indent With Theorem 2, we indicate that function $Sum(.)$ is submodular.
\begin{theorem} 
\textit{For an arbitrary instance of the diversity maximization problem, the resulting optimization function $Sum(.)$ is submodular.}
\end{theorem}
\begin{proof}
\textit{In order to establish this result, we need to prove that $\forall C, w_0, w_1$, we have
\begin{equation}
\begin{aligned}
    Sum(C\cup \{w_0\}) + Sum(C\cup \{w_1\}) \geq\\ Sum(C\cup \{w_0,w_1\}) + Sum(C) 
\end{aligned}
\end{equation}
where $C\subseteq W$, $w_0,w_1 \in W - C$. By Definition 2, we express the left-hand-side and right-hand-side as follows
\begin{equation}
\begin{aligned}
    LHS = &- \sum_{w_i,w_j\in C \wedge i\neq j}Sim(w_i,w_j) - \sum_{w\in C}Sim(w,w_0)\\ &- \sum_{w_i,w_j\in C \wedge i\neq j}Sim(w_i,w_j) - \sum_{w\in C}Sim(w,w_1)
\end{aligned}
\end{equation}
\begin{equation}
\begin{aligned}
    RHS = &- \sum_{w_i,w_j\in C \wedge i\neq j}Sim(w_i,w_j) - \sum_{w\in C}Sim(w,w_0)\\& - \sum_{w_i,w_j\in C \wedge i\neq j}Sim(w_i,w_j) - \sum_{w\in C}Sim(w,w_1) \\&- Sim(w_0,w_1)
\end{aligned}
\end{equation}
\indent Therefore, we have:
\begin{equation}
    LHS - RHS = Sim(w_0,w_1) \geq 0
\end{equation}
which competes the proof.} \hfill $\square$
\end{proof}

\indent Facilitated by Theorem 2, our first main result is that the optimal solution for \textit{diversity maximization} can be efficiently approximated within a factor of $(1 - 1/e - \epsilon)$ \cite{Ref7}. Here $e$ is the base of the natural logarithm and $\epsilon$ is any arbitrary small positive real number. Thus, this is a performance guarantee slightly better than $(1 - 1/e) = 63\%$.\\
\indent Algorithm 1 lists the detailed steps of this approximation algorithm. This algorithm, which achieves the performance guarantee, is a natural greedy hill-climbing strategy related to the approach considered in \cite{Ref7}. Thus the main content of this result is the analysis framework needed for obtaining a provable performance guarantee, and the fairly surprising fact that hill-climbing is always within a factor of at least 63\% of the optimal for this problem.\\
\begin{algorithm}[ht]
    \SetAlgoLined
    \SetKwInOut{Input}{Input}
    \SetKwInOut{Output}{Output}
    \Input{$C\leftarrow \emptyset$}
    \Output{Find $C$ s.t. $|C|$ $= k$ and $Div(C)$ is maximized.}
    $C\leftarrow\{w_0,w_1\}$\;
    \While{$|C|\leq k$}{
    \begin{equation}\nonumber
        \quad \quad x = \mathop{\arg\max}_{w_x\in W} Div(C\cup\{w_x\})\;
    \end{equation}
    $C\leftarrow C\cup\{w_x\}$\;
    }
    \Return C\;
    \caption{Diversity Maximization}
\end{algorithm}
\indent In Algorithm 1, the selection of $w_0$ and $w_1$ can vary, leading to divergent outcomes. Considering the experimental findings and acknowledging the limitations associated with local optimal solutions, we adopt two distinct selection methods in this study. Firstly, we opt for selecting workers with the smallest $Sim$  value as $w_0$ and $w_1$ (Here, $Div(C)=-Sim_{w_0,w_1}/2$), aligning with the inherent logic of the greedy algorithm itself (referred to as `MIN-SIM'). Secondly, we choose workers with the smallest sum of related $Sim$ values as $w_0$ and $w_1$ (i.e., $\arg\min_{i,j}\sum_{k}(Sim_{i,k}+Sim_{j,k})$, referred to as `MIN-SUM'). This adjustment accounts for extreme scenarios. We will delineate the advantages and drawbacks of these two initialization methods in the experiments section for reference.
\section{TASK-DRIVEN MODEL}
\label{T-model part}
Under the task-driven model, each worker is associated with a probability, describing his/her opinion about the given task. We aim to select $k$ workers out of $n$ candidates, such that the numbers of positive and negative workers satisfy a user’s demand.\\
\indent We formally define the optimization problem and related important notations in this section.
\begin{definition}[WORKER OPINION] \textit{A crowdsourcing worker $w_i$ is associated with an opinion $t_i$ about the given task, which is a Bernoulli random variable. We denote the probability $Pr(t_i = 1) = 1-Pr(t_i = 0)$, where $Pr(t_i = 1)(Pr(t_i = 0))$ is the probability of $w_i$ having a positive(negative) opinion about the task. We assume that the opinions of all the workers are independent.}\\\\
\end{definition}
\indent There are two possible ways to obtain the probabilities for the workers. Firstly, when a crowdsourcing platform is implemented on a public online community (e.g. social networks, online forums), we can analyze the historical data and profile information of a given user. Any of the current techniques can be used as a plug-in for our system to detect relevance of a worker to a subject of interest. Secondly, before selecting a worker to participate in a crowd, we may simply ask individual workers for their opinions towards the given subject. On common crowdsourcing platforms, such questions can be designed as so-called \emph{Qualification Tests}, which are prerequisites for workers to answer any questions thereafter.
\subsection{Crowd Selection with T-model}
\label{The principle of T model}
Now we illustrate how to optimize the process of worker selection under T-model. Before providing the formal definition, we introduce the rationale of the optimization. Since each worker’s opinion is probabilistic, the total number of workers with positive (negative) opinions is also a probabilistic distribution. We assume that we have the user’s demand of the number of workers with positive (negative) opinions, and the optimization is to select the best subset of workers such that the user’s demand is satisfied.\\
\indent As follows, we define the optimization problem under T-model.
\begin{definition}[K-BEST WORKERS SELECTION]\textit{Given a set of $|N|$ workers $w_i,w_2,\cdot\cdot\cdot,w_{|N|}$ with opinions $N = \{t_1,t_2,\cdot\cdot\cdot,t_{|N|}\}$. Let $\theta_1$ and $\theta_0$ be the user’s demand on the numbers of workers being supportive or opposing with respect to the given task, respectively. We aim to select $k$ workers, so that the probability of the user’s demand being fulfilled is maximized. To ensure this probability is positive for any $k \geq 1$, we assume $\theta_0+\theta_1 \leq k$. Formally, let $S$ be the subset of $N$, and let $\tau$ be the probability that at least $\theta_1$($\theta_0$) workers existing in $S$ supporting (opposing) the given task,}
\begin{equation}
    \tau(S) = Pr\{\sum_{t\in S}t\geq \theta_1 \wedge \sum_{t\in S}(1 - t) \geq \theta_0\}
\end{equation}
\textit{we have the optimization problem as follows:}
\begin{equation}
    S:= \mathop{\arg\max}_{|S|=k} \tau(S)\
\end{equation}
\end{definition}
\indent By taking a closer look at Formula 18, we have $\sum_{t\in S}t+\sum_{t\in S}(1-t)=k$. For the sake of presentation, we denote $T = \sum_{t\in S}t, \theta_2 = k - \theta_0$. Then, Formula 18 can be rewritten as
\begin{equation}
\begin{aligned}
    \tau(S) &= Pr(\theta_1 \leq T \leq \theta_2)\\
    & = \sum_{i=\theta_1}^{\theta_2}Pr(T=i)
\end{aligned}
\end{equation}
Since each worker can be treated as a random variable following Bernoulli distributions, $T$ follows a standard Poison Binomial distribution (PBD). Therefore, by adopting the probability mass function (pmf) of PBD, we have
\begin{equation}
    \tau(S) = \sum_{i=\theta_1}^{\theta_2}\sum_{A\in F_t}\prod_{t_\alpha\in A}Pr(t_\alpha=1)\prod_{t_\beta\in A^c}Pr(t_\beta=0)
\end{equation}
where $F_t$ is the set of all the subsets of $S$.
\begin{example} A concrete example to illustrate the optimization problem as illustrated in Fig.\ref{fig:2}. Assume we have a set of candidate workers, with worker opinions 0.2, 0.3, 0.4, 0.6, 0.8 and 0.9, respectively. We further assume that a user wants to select 4 of them, and one of them has a positive opinion and one of them has a negative opinion. Hence, we have $\theta_1 = 1,\theta_0 = 1,k = 4$, then $\theta_2 = 4 - 1 = 3$. There are totally $C_6^4$ possible combinations, each of which indicates a PBD. We present all the possible size-4 combinations, and compute $\tau(S)$ for each of them in Table 3. Fig.\ref{fig:3} illustrates the PBD of the number of workers with positive opinions, and indicates the range of probabilities we aim to maximize.
\begin{figure}
\centering
  \includegraphics[scale=0.35]{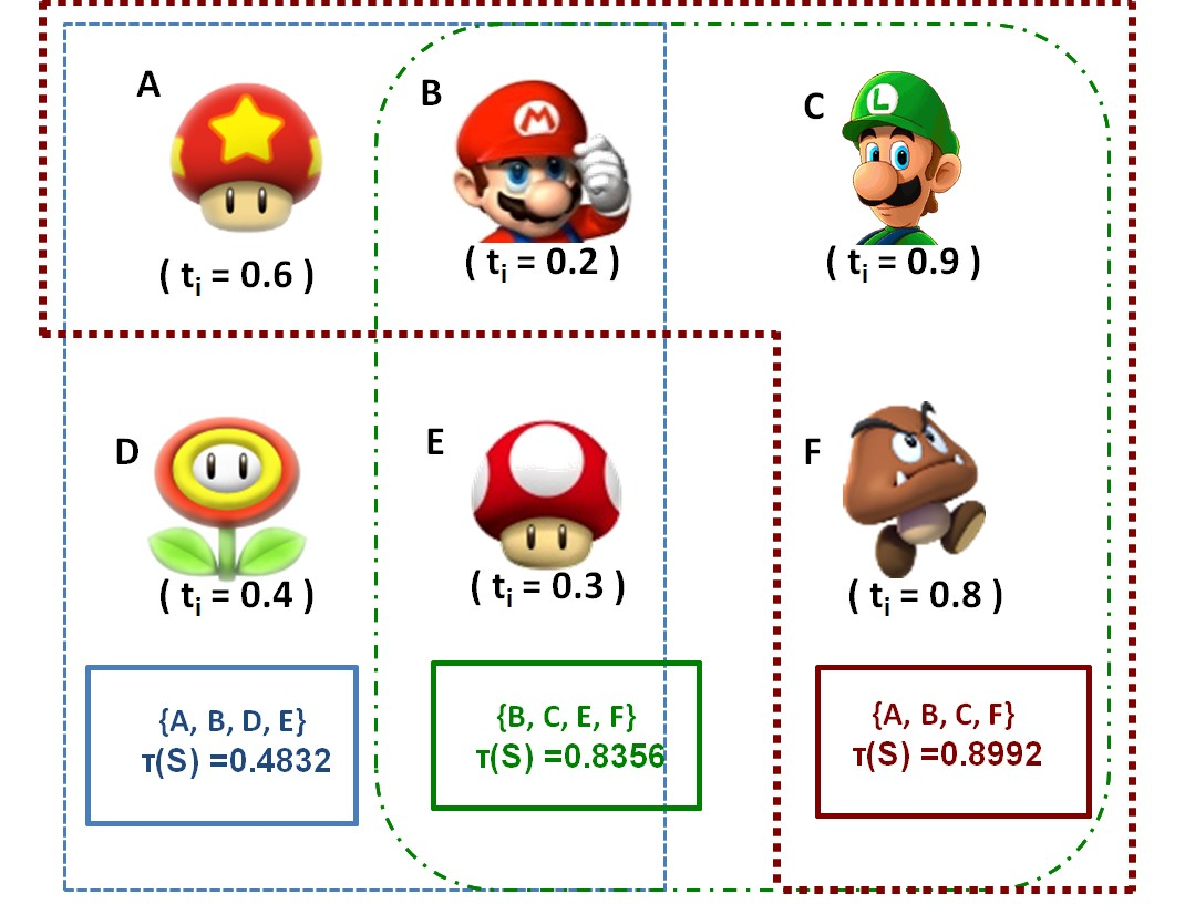}
\caption{Find 4 workers including 1 supporter and 1 objector. The values of $t_{i}$'s represent worker's opinions, and the three rectangular represent three concrete cases of how to compute $\tau(S)$ for four workers with 1 supporter and 1 objector based on the PBD.}
\label{fig:2}       
\end{figure}
\begin{table}[ht]\centering
\begin{tabular}{|c|c|c|c|}
\hline
Crowd & $\tau(S)$ & Crowd & $\tau(S)$\\ \hline
A,B,C,D & 0.7616 & A,B,C,E & 0.7272 \\ \hline
A,B,C,F & 0.8992 & A,B,D,E & 0.4832 \\ \hline
A,B,D,F & 0.7152 & A,B,E,F & 0.6784 \\ \hline
A,C,D,E & 0.7884 & \textbf{A,C,D,F} & \textbf{0.9224} \\ \hline
A,C,E,F & 0.9108 & A,D,E,F & 0.7448 \\ \hline
B,C,D,E & 0.6188 & B,C,D,F & 0.8568 \\ \hline
B,C,E,F & 0.8356 & B,D,E,F & 0.5736 \\ \hline
C,D,E,F & 0.8732 & & \\ \hline
\end{tabular}
\caption{$\tau(S)$ values of all 4-worker combinations in Fig.\ref{fig:2}}
\end{table}
\begin{figure}[ht]\centering
\centering
  \includegraphics[scale=0.35]{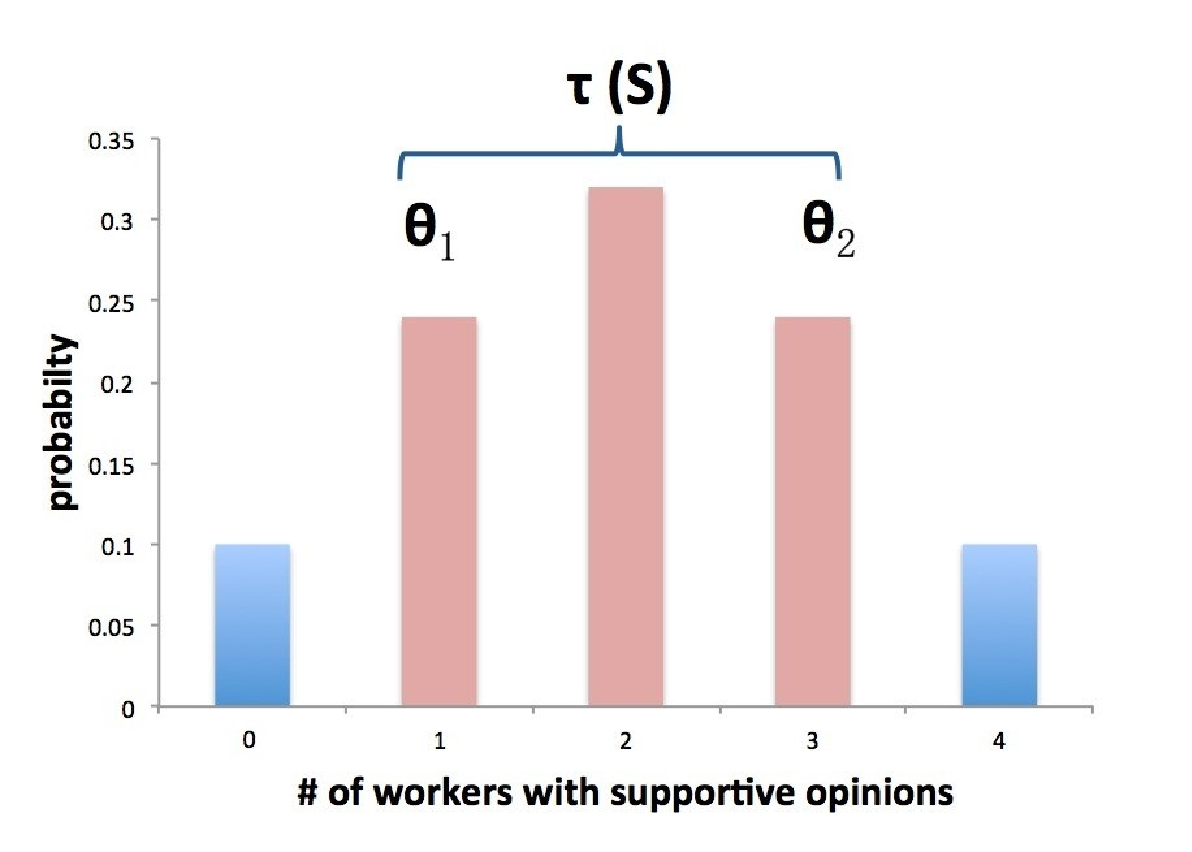}
\caption{The Poisson-Binomial Distribution}
\label{fig:3}       
\end{figure}

\indent One can see that $< A,C,D,F >$ is the optimal choice, since it maximizes the probability that the user’s demand is satisfied.
\end{example}
\subsection{Method with Poisson Approximation}
\label{Poisson Approximation}
To select the exact optimal combination of $k$ workers, we have to enumerate all $O(n^k)$ PBDs, and output the one with the highest $\tau(S)$. However, this naive method leads to very high computational cost. In this subsection, we consider each PBD as a Poisson distribution, and conduct the selection among the approximated Poisson distributions. By aborting the bounded imprecision introduced by the approximation, we significantly improve the efficiency.\\
\indent A Poisson Binomial distribution can be well approximated by a Poisson distribution. Then, we consider $T$ approximately following a Poisson distribution, with parameter $\lambda = \sum_{t\in S}Pr(t=1)$.\\
\indent Then we have
\begin{equation}
    Pr(\theta_1 \leq T \leq \theta_2) \approx F_P (\theta_2,\lambda) - F_P(\theta_1,\lambda)
\end{equation}
where $F_P$ is the cumulative mass function (CMF) of the Poisson distribution. As a result, we find $S'$ to maximize
\begin{equation}
    G_P(\lambda):= F_P (\theta_2,\lambda) - F_P (\theta_1,\lambda)
\end{equation}
and return $S'$ as the approximate answer. In the reminder of this subsection, we first analyze the monotonicity of $G_P(\lambda)$, and then provide two algorithmic solutions.
\subsubsection{Monotonicity Analysis}
\label{Poisson distribution's monotonicity analysis}
In the following, we first analyze the monotonicity of $G_P(\lambda)$. We discover that $G_P(\lambda)$ has a nice monotonic property, which is algorithmically useful. This discovery is concluded with the following theorem.
\begin{theorem}
\textit{Considering $\lambda$ as a continues independent variable with range $(0,k)$, $G_P(\lambda)$ monotonously increases and decreases on $[0,(\frac{\theta_2!}{\theta_1!})^{\frac{1}{\theta_2-\theta_1}}]$ and $[(\frac{\theta_2!}{\theta_1!})^{\frac{1}{\theta_2-\theta_1}}, k]$, respectively}.
\end{theorem}
\begin{proof} 
First, we expand $F_P$, the CMF of Poisson distribution, and rewrite $G_P(\lambda)$ as
\begin{equation}
\begin{aligned}
    G_P(\lambda) & = e^{-\lambda}\sum_{i=0}^{\theta_2}\frac{\lambda^i}{i!} - e^{-\lambda}\sum_{j=0}^{\theta_1}\frac{\lambda^j}{j!}\\
    & =\sum_{i=\theta_1+1}^{\theta_2}\frac{e^{-\lambda}\lambda^i}{i!}
\end{aligned}
\end{equation}
Then, we take the partial derivative of $G_P(\lambda)$ w.r.t $\lambda$:
\begin{equation}
\begin{aligned}
    \frac{\partial G_P(\lambda)}{\partial\lambda} &= \sum_{i=\theta_1+1}^{\theta_2}\frac{\partial\frac{e^{-\lambda}\lambda^i}{i!}}{\partial\lambda} \\&= \sum_{i=\theta_1+1}^{\theta_2}\frac{e^{-\lambda}(i\lambda^{i-1}-\lambda^i)}{i!} \\ 
    &=e^{-\lambda}\sum_{i=\theta_1+1}^{\theta_2}\frac{(i\lambda^{i-1}-\lambda^i)}{i!} \\&= e^{-\lambda}\sum_{i=\theta_1+1}^{\theta_2}\{\frac{\lambda^{i-1}}{(i-1)!}-\frac{\lambda^i}{i!}\}\\
    &=e^{-\lambda}\{\sum_{i=\theta_1+1}^{\theta_2}\frac{\lambda^{i-1}}{(i-1)!}-\sum_{i=\theta_1+1}^{\theta_2}\frac{\lambda^i}{i!}\} \\&= e^{-\lambda}\{\frac{\lambda^{\theta_1}}{\theta_1!}-\frac{\lambda^{\theta_2}}{\theta_2!}\}\\
    &=e^{-\lambda}\lambda^{\theta_1}\{\frac{1}{\theta_1!}-\frac{\lambda^{\theta_2-\theta_1}}{{\theta_2!}}\}
\end{aligned}
\end{equation}
To analyze the monotonicity of $G_P(\lambda)$, we solve $\lambda$ for inequation $\frac{\partial G_P(\lambda)}{\partial\lambda} > 0$. Note that, in eq. 25, we have $e^{-\lambda}\lambda^{\theta_1} > 0$, and $\theta_2 > \theta_1$, so
\begin{equation}
\begin{aligned}
    \frac{\partial G_P(\lambda)}{\partial\lambda}=e^{-\lambda}\lambda^{\theta_1}\{\frac{1}{\theta_1!}-\frac{\lambda^{\theta_2-\theta_1}}{\theta_2!}\}>0\\
    \Leftrightarrow\lambda^{\theta_2-\theta_1}<\frac{\theta_2!}{\theta_1!}\Leftrightarrow\lambda<(\frac{\theta_2!}{\theta_1!})^{\frac{1}{\theta_2-\theta_1}}
\end{aligned}
\end{equation}
Similarly, we have $\frac{\partial G_P(\lambda)}{\partial\lambda} < 0 \Leftrightarrow \lambda>(\frac{\theta_2!}{\theta_1!})^{\frac{1}{\theta_2-\theta_1}}$, which completes the proof. \hfill $\square$
\end{proof}
\subsubsection{Transformation to Exact k-item Knapsack Problem (E-kKP)}
\label{E-kKP algorithm intro}
Based on the discovered monotonicity property, we show that maximizing $G(\lambda)$ is equivalent to the classical “Exact k-object Knapsack (E-kKP)” problem as shown by the following Theorem.
\begin{theorem}
\textit{By considering each PBD approximately as a Poisson distribution, the k-best workers selection problem can be solved by any algorithm for the Exact k-item Knapsack Problem (E-kKP).}
\end{theorem}
\begin{proof}
\textit{Facilitated with theorem 3, our optimization is revised to select $S$ such that $\lambda = \sum_{t\in S}Pr(t = 1)$ approaches
$(\frac{\theta_2!}{\theta_1!})^{\frac{1}{\theta_2-\theta_1}}$, which is a constant number. Furthermore, we have $\lambda = \sum_{t\in S}Pr(t = 1)$, then by defining}
\begin{equation}
    \Omega_P:=(\frac{\theta_2!}{\theta_1!})^{\frac{1}{\theta_2-\theta_1}}
\end{equation}
\textit{our optimization is further revised as selecting $S$ such that $\sum_{t\in S}Pr(t = 1)$ approaches $\Omega_P$. Despite having the nice property of monotonicity, $G_P(\lambda)$ may not be symmetric, and $\lambda=\sum_{t\in S}Pr(t = 1)$ is a discrete variable. This indicates, we need to find $\lambda_l$ and $\lambda_r$, which achieve maximums of $G_P$ on $[0,(\frac{\theta_2!}{\theta_1!})^{\frac{1}{\theta_2-\theta_1}}]$ and $[(\frac{\theta_2!}{\theta_1!})^{\frac{1}{\theta_2-\theta_1}}$ $,k]$, respectively. Then we choose between them by comparing $G_P(\lambda_l)$ and $G_P(\lambda_r)$. Consequently, we aim to find two size-k subsets $S_l$ and $S_r$ of the given $N$, such that $\sum_{t\in S_l}Pr(t = 1)$ is largest to but no larger than $\Omega_P$ , and $\sum_{t\in S_r}Pr(t = 1)$ is smallest to but smaller than $\Omega_P$. Actually, algorithmically speaking, finding $S_l$ is the same as finding $S_r$. This is because finding $S_r$ is equivalent to finding $N-S_r$, which is $|N|-k$ sized, such that $\sum_{t\in N-S_r}Pr(t = 1)$ is the largest but no larger than $\sum_{t\in N}Pr(t = 1)-\Omega_P$. Therefore, the remaining optimization problem is: finding $S_l$, which is a size-k subset of N, and we want to maximize the sum of values in $S_l$ without exceeding $\Omega_P$. This is a typical E-kKP problem.} \hfill $\square$
\end{proof}
It is known that E-kKP can be solved by\\
\indent(1) a backtracking approach with $O(|N|^k/k!)$ time;\\
\indent(2) dynamic programming with $O(\gamma|N|)$;\\
\indent(3) 1/2-approximation algorithm by linear programming with $O(|N|)$.\\
\indent These three algorithms are proposed in \cite{Ref11}. For showing how to adopt these algorithms, we only demonstrate (1), that is, the backtracking algorithm with Algorithm 2. The other two algorithms are analogous.\\
\begin{algorithm}
\SetAlgoLined
\SetKwInOut{Input}{Input}
\SetKwInOut{Output}{Output}
\Input{$k,\Omega,N=\{t_1,...,t_{|N|}\}$}
\Output{A size-k subset of $N$}
Function $Bt(k,\Omega,N)$\;
\uIf{$|N|=k$}{
\Return $N$\; 
}
\uElseIf{$\sum_{i=1}^{k}Pr(t_i=1)>\Omega$}
{\Return null\;}
\uElseIf{$Bt(k,\Omega,N-t_{|N|}) > Bt(k-1,\Omega-Pr(t_{|N|}=1),N-t_{|N|})+Pr(t_{|N|}=1)$}
{\Return $Bt(k,\Omega,N-t{|N|})$\;}
\Else{\Return $Bt(k-1,\Omega-Pr(t_{|N|}=1),N-t_{|N|})\cup t_{|N|}$\;}
\caption{Backtracking Algorithm (Bt)}
\end{algorithm}
\indent With Algorithm 2, we find $S_l$ and $S_r$ by $Bt(k,\Omega_P,N)$ and $N - Bt(|N|- k,\sum_{t\in N}Pr(t=1)-\Omega_P,N)$, receptively. Note $\lambda_l = \sum_{S_l}Pr(t=1)$ and $\lambda_r = \sum_{S_r}Pr(t = 1)$, we set the output $S^{\prime} = S_l$ as the final result if $G(\lambda_l) > G(\lambda_r)$; otherwise $S^{\prime} = S_r$ is returned.
\subsection{Method with Binomial Approximation}
\label{Binomial Approximation}
It is known that Binomial approximation is also an effective method to deal with the high complexity of the Poisson Binomial distribution. Similar to the Poisson approximation, we have
\begin{equation}
    Pr(\theta_1 \leq T \leq \theta_2) \approx F_B (\theta_2;n,p) - F_B(\theta_1;n,p)
\end{equation}
where $F_B$ is the CMF of Binomial distribution with parameter $n = k $ and $p=\frac{\sum_{t\in S}Pr(t=1)}{k}$ . Then, the optimization is to maximize:
\begin{equation}
    G_B(P) :=  F_B (\theta_2;n,p) - F_B(\theta_1;n,p)
\end{equation}
Please note $n$ is a fixed parameter since $k$ is a constant in K-best workers selection problem. Therefore, what we can do is to simply adjust $p$ with different selections of $S$. Analogous to the Poisson approximation in Section \ref{Poisson Approximation}, we first analyze the monotonicity, and then discuss the algorithm.\\
\indent\textbf{Monotonicity Analysis}: With theorem 5, we show that $G_B(p)$ also has a useful monotonic feature, which is similar to the Poisson approximation.
\begin{theorem}\textit{Considering $p$ as a continues independent variable with range $(0,n)$, $G_B(p)$ monotonously increases and decreases on $[0,\frac{1}{1+(\frac{(n-\theta_2)C_n^{\theta_2}}{(n-\theta_1)C_n^{\theta_1}})^{\frac{1}{\theta_2-\theta_1}}}]$ and $[\frac{1}{1+(\frac{(n-\theta_2)C_n^{\theta_2}}{(n-\theta_1)C_n^{\theta_1}})^{\frac{1}{\theta_2-\theta_1}}}$ $,n]$, respectively.}
\end{theorem}
\begin{proof}
\textit{The CMF of a Binomial distribution, $F_B$, can be represented in terms of the regularized incomplete beta function:
\begin{equation}
    F_B(\theta;n,p) = (n-\theta)C_n^\theta \int_0^{1-p}t^{n-\theta-1}(1-t)^\theta dt
\end{equation}
Facilitated with Formula 30, we compute the partial derivative of
$G_B(p)$ w.r.t $p$:
\begin{equation}
\begin{aligned}
    \frac{\partial G_B(p)}{\partial p}&=(n-\theta_2)C_n^{\theta_2}\frac{\partial\int_0^{1-p}t^{n-\theta_2-1}(1-t)^{\theta_2}dt}{\partial p}\\&\quad-(n-\theta_1)C_n^{\theta_1}\frac{\partial\int_0^{1-p}t^{n-\theta_1-1}(1-t)^{\theta_1}dt}{\partial p}\\&=(n-\theta_2)C_n^{\theta_2}\{-(1-p)^{n-\theta_2-1}p^{\theta_2}\}\\&\quad-(n-\theta_1)C_n^{\theta_1}\{-(1-p)^{n-\theta_1-1}p^{\theta_1}\}\\&=p^{\theta_1}(1-p)^{n-\theta_2-1}\{(n-\theta_1)C_n^{\theta_1}(1-p)^{\theta_2-\theta_1}\\&\quad-(n-\theta_2)C_n^{\theta_2}p^{\theta_2-\theta_1}\}
\end{aligned}
\end{equation}
Then, by solving equations $\frac{\partial G_B(p)}{\partial p} \geq 0$ and $\frac{\partial G_B(p)}{\partial p} \leq 0$, we have results $p \leq \frac{1}{1+(\frac{(n-\theta_2)C_n^{\theta_2}}{(n-\theta_1)C_n^{\theta_1}})^{\frac{1}{\theta_2-\theta_1}}}$ and $p \geq \frac{1}{1+(\frac{(n-\theta_2)C_n^{\theta_2}}{(n-\theta_1)C_n^{\theta_1}})^{\frac{1}{\theta_2-\theta_1}}}$, respectively, which completes the proof. \hfill $\square$}
\end{proof}
\indent \textbf{$\quad$Algorithms}: Algorithm 2 (and other algorithms for E-kKP problem) can be reused for finding the approximate solution based on Binomial approximation. Specifically, we define
\begin{equation}
    \Omega_B:= \frac{k
    }{1+(\frac{(n-\theta_2)C_n^{\theta_2}}{(n-\theta_1)C_n^{\theta_1}})^{\frac{1}{\theta_2-\theta_1}}}
\end{equation}
and the solution subset is between $S_l^{\prime} = Bt(k,\Omega_B,N)$ and $S_r^{\prime} = N - Bt(|N|-k,\sum_{t\in N}Pr(t=1)-\Omega_B,N)$. Here, let $p_l = \sum_{t\in S_l^{\prime}}Pr(t=1)$ and $p_r = \sum_{t\in S_r^{\prime}}Pr(t=1)$, then we return $S_l^{\prime}$ as result if $G_B(p_l) > G_B(p_r)$; otherwise return $S_r^{\prime}$.
\subsection{Method with Normal Approximation}
\label{Normal Approximation and Simulated Annealing method}
It's well-known that both the Poisson distribution and Binomial distribution can be approximated by normal distribution. In previous sections, we have already discussed methods using Poission and Binomial distributions to approximate the Poisson Binomial distribution. Now we will further use normal distribution to approximate the Poisson Binomial distribution. We assume that $T$ approximately obeys a normal distribution with the expection:
\begin{equation}
    \mu_S = \sum_{t\in S}{Pr(t=1)}
\end{equation}
and the standard deviation:
\begin{equation}
    \sigma_S = \sqrt{\sum_{t\in S}{(Pr(t=1)(1-Pr(t=1)))}}
\end{equation}
Then, Formula 18 can be approximated as:
\begin{equation}
\begin{aligned}
 \tau(S) &= Pr(\theta_1 \leq T \leq \theta_2) \\
 &\approx F_N(\theta_2+0.5;\mu_S,\sigma_S)  - F_N(\theta_1-0.5;\mu_S,\sigma_S)
\end{aligned}
\end{equation}
where $F_N$ is the CDF of Normal distribution with parameters $\mu_S$ and $\sigma_S$. Here we add a continuity correction according to \cite{Ref37}, considering that the Poisson Binomial distribution is discrete while the normal distribution is continuous. Now, we find $S^{\prime}$ to maximize:
\begin{equation}
    G_N(\mu_S,\sigma_S) := F_N(\theta_2+0.5;\mu_S,\sigma_S)  - F_N(\theta_1-0.5;\mu_S,\sigma_S)
\end{equation}
\indent Unlike the Poisson approximation and Binomial approximation, $G_N(\mu_S,\sigma_S)$ has no monotonic features, and it contains two independent parameters ($\mu_S$ and $\sigma_S$) required to be controlled. Therefore, it's not suitable to utilize only one parameter ($\mu_S$ or $\sigma_S$) to help us finding the best group by the aforementioned Knapsack method (Algorithm 2). If we utilize the two parameters simultaneously, the time cost by normal approximation would be much higher than that by the Poisson or Binomial approximation. Thus, it is necessary to choose an algorithm that can not only take advantage of fast computation of $G_N(\mu_S, \sigma_S)$ but also decrease the time cost.\\
\indent The simulated annealing algorithm \cite{Ref38} is a good choice to deal with the issue mentioned above. It is a stochastic optimization algorithm based on the Monte-Carlo iterative solution strategy, and its starting point is based on the similarity between the annealing process of solid matter in physics and general combinatorial optimization problems. The simulated annealing algorithm starts from a relatively high initial temperature, and with the continuous decrease of the temperature parameter, combined with a certain probability jump characteristic, randomly finds the global optimal solution of the objective function in the solution space, that is, the local optimal solution can be probabilistically jump out and eventually tend to the global optimum.\\
\indent We demonstrate the algorithm in Algorithm 3, where $T_{ini}$, $T_{end}$, $r$ and $c$ represent the initial temperature, the terminal temperature, repeating times and the ratio of chilling-down, respectively. Similar to the original algorithm which improves the results by adding random displacement to the atom at each iteration, our algorithm improves the results by randomly replacing a small number ($k_1$) of elements in $S$ with $k_1$ elements from its complementary set $R$ as shown in Algorithm 3.\\
\begin{algorithm}[ht]
    \SetAlgoLined
    \SetKwInOut{Input}{Input}
    \SetKwInOut{Output}{Output}
    \Input{$k, N=\{t_0,t_1,...,t_{N-1}$\}}
    \Output{A size-k subset of $N$}
    Initialize $T=T_{ini}$, $r$, $c$\;
    $S$ $\leftarrow$ A set consisting of random $k$ elements in $N$\;
    $R$ $\leftarrow$ $N - S$\;
    \While{$T > T_{end}$}{
    \For{$t \leftarrow 0$ \KwTo $r$}{
    $k_1$ $\leftarrow$ random(1, floor(min($\frac{k}{2}$,$\frac{|N|-k}{2}$)))\;
    $R_1$ $\leftarrow$ A set consisting of random $k_1$ elements in $R$\;
    $S_1$ $\leftarrow$ A set consisting of random $k-k_1$ elements in $S$\;
    $S_1$ $\leftarrow$ $S_1 \bigcup R_1$\;
    $\Delta G_N \leftarrow G_N(\mu_{S_1},\sigma_{S_1}) - G_N(\mu_S,\sigma_S)$\;
    \eIf{$\Delta G_N > 0$}{
        $S \leftarrow S_1$\;
        $R \leftarrow N - S_1$\;
        }{
        Accept $S_1$ with probability $e^{\frac{\Delta G_N}{T}}$\;
        \If{Accept $S_1$}{
        $S \leftarrow S_1$\;
        $R \leftarrow N - S_1$\;}
        }
        }
        $T \leftarrow T*c$\;
    }
    \Return $S$\;
    \caption{Simulated Annealing Algorithm}
\end{algorithm}
\indent The computation of $\tau(S)$ will be the main factor of affecting the time complexity of the algorithm. We use the normal approximation to calculate $\tau(S)$ (eq. 35). Suppose the time complexity of calculating $\tau(S)$ is $k$, then the total time complexity of this algorithm is $O(kr\log_{c}(\frac{T_{ini}}{T_{end}}))$. It relates to the hyper-parameters $T_{ini}$, $T_{end}$, $r$ and $c$. All of them are chosen by the user and have no connection to $k$ and $|N|$. Users could balance the time complexity and the accuracy they want by selecting the suitable parameters. In the experiments of Section \ref{Experiments on synthetic data}, we set the parameters as $T_{ini}=1$, $T_{end}=0.0001$, $r=1000$, $c=0.9$.\\
\indent Compared to the aforementioned Poisson approximation and Binomial approximation that approximate $\tau(S)$ twice, the normal approximation with the simulated annealing algorithm only approximates $\tau(S)$ once. Thus, we believe the latter can improve the results, which is demonstrated by the experimental evaluation in Section \ref{Experiments on T-model}.
\subsection{Method with Discret Fourier Transform and Character Function (DFT-CF)}
\label{exact DFT-CF method}
The DFT-CF method is an exact method of computing the probability mass function \cite{Ref37,Ref38}. Here, it is based on two-dimensional Discret Fourier Transform (DFT) of the Character Function (CF) of the Poisson Binomial distribution.\\
\indent The Character Function of the Poisson Binomial variable $T=\sum_{j}t_j$ is given by $\phi(t) = E(e^{\boldsymbol{i}tT})$. According to the definition of mathematical expectation, we have:
\begin{equation}
    \phi(t) = \sum_{T_0=0}^{k}Pr(T=T_{0})e^{itT_0}.
\end{equation}
On the other hand, we assume that $t_j$'s are mutually independent, so
\begin{equation}
    \phi(t) = \prod_{j=1}^{k}E(e^{\boldsymbol{i}tt_j}) = \prod_{j=1}^{k}(1+Pr(t_j=1)(e^{\boldsymbol{i}t}-1))
\end{equation}
Let $t=\omega l$, where $l$ is 0...$k$, $\omega = 2\pi/(k+1)$, by combining equations (37) and (38), we have:
\begin{equation}
\begin{aligned}
    \frac{1}{k+1}\sum_{T_0=0}^{k}{Pr(T=T_0)e^{\boldsymbol{i}\omega lT_0}} =\\
    \frac{\prod_{j=1}^{k}{(1+Pr(t_j=1)(e^{\boldsymbol{i}\omega l}-1))}}{k+1}
\end{aligned}
\end{equation}
Application of DFT to both sides of eq. (39) leads to the pmf of T as follows (see details in \cite{Ref38}):
\begin{equation}
    Pr(T=T_0) = \frac{1}{k+1}\sum_{l=0}^k{(\prod_{j=1}^k{((1-p_j)+p_je^{\boldsymbol{i}\omega l}})e^{-\boldsymbol{i}\omega lT_0})}
\end{equation}
Thus, we have:\\ 
\begin{equation}
\begin{aligned}
    \tau(S) &= Pr(\theta_1 \leq T \leq \theta_2) \\&= \frac{1}{k+1}\sum_{T_0=\theta_1}^{\theta_2}\sum_{l=0}^k{(\prod_{j=1}^k{((1-p_j)+p_je^{\boldsymbol{i}\omega l}})e^{-\boldsymbol{i}\omega lT_0})}
\end{aligned}
\end{equation}
\indent Now we can still utilize the simulated annealing algorithm by replacing the $G_N(\mu_S, \sigma_S)$ in Algorithm 3 with the exact $\tau(S)$ in eq. (41).\\
\indent The time complexity of calculating $Pr(T=T_0)$ is $O(k^2)$, where $k$ is the size of $S$. Thus, it is straightforward to get the time complexity of the method DFT-CF (see $\tau(S)$ in eq. 41) is $O(k^2(\theta_2-\theta_1))$. Compared to the normal approximation in Section \ref{Normal Approximation and Simulated Annealing method}, DFT-CF needs not to approximate $\tau(S)$. Therefore, the results by DFT-CF tend to be more accurate. The experimental evaluation in Section \ref{Experiments on T-model} demonstrate this. 
\section{EXPERIMENTAL EVALUATION}
\label{Expermental evaluation and Case study}
In this section, we present our experimental evaluation of the performances of T-model and S-model, as well as an experimental study of the crowd selection problem, namely finding the optimal set of workers with a given budget. The goal of our experiments is twofold: first, we study the effect of different parameters for the proposed algorithms; second, we compare the two proposed algorithms with a baseline algorithm, that is, selecting the workers randomly. In order to explore the various settings of parameter values in our methods, we have used synthetic data for the testing. In addition, we verify the effectiveness of our methods on data from the Foursquare \cite{Ref1}, a very popular social network. Specifically, we used the Foursquare API to gather sample data of the existing venues and the tips posted on them. In particular, for each collected venue, the crawler collects all its tips, the identifications of the users who posted each of them. Our crawler ran from March 15th to May 19th, which collected data from 69,423 users. We also verify the effectiveness of S-model on the Music \& Mental Health Survey Results dataset \cite{Ref53} obtained from Kaggle. This dataset consists of 618 complete survey responses containing personal information and music preferences, with over 30 questions. Additionally, to evaluate the practicability of the proposed models, we conducted a case study on Amazon Mechanical Turk (AMT).\\ 
\indent All the experiments are conducted on a server equipped with Intel(R) Core(TM)i7 3.40GHz PC and 16GB memory, running on Microsoft Windows 7.
\subsection{Experiments on S-model}
\label{Experiments on S-model}
\subsubsection{Synthetic data}
\label{Synthetic data experiments on S-model}
\begin{figure*}[h]
\centering
  \subfloat[Uniform Distribution]{\includegraphics[scale=0.4]{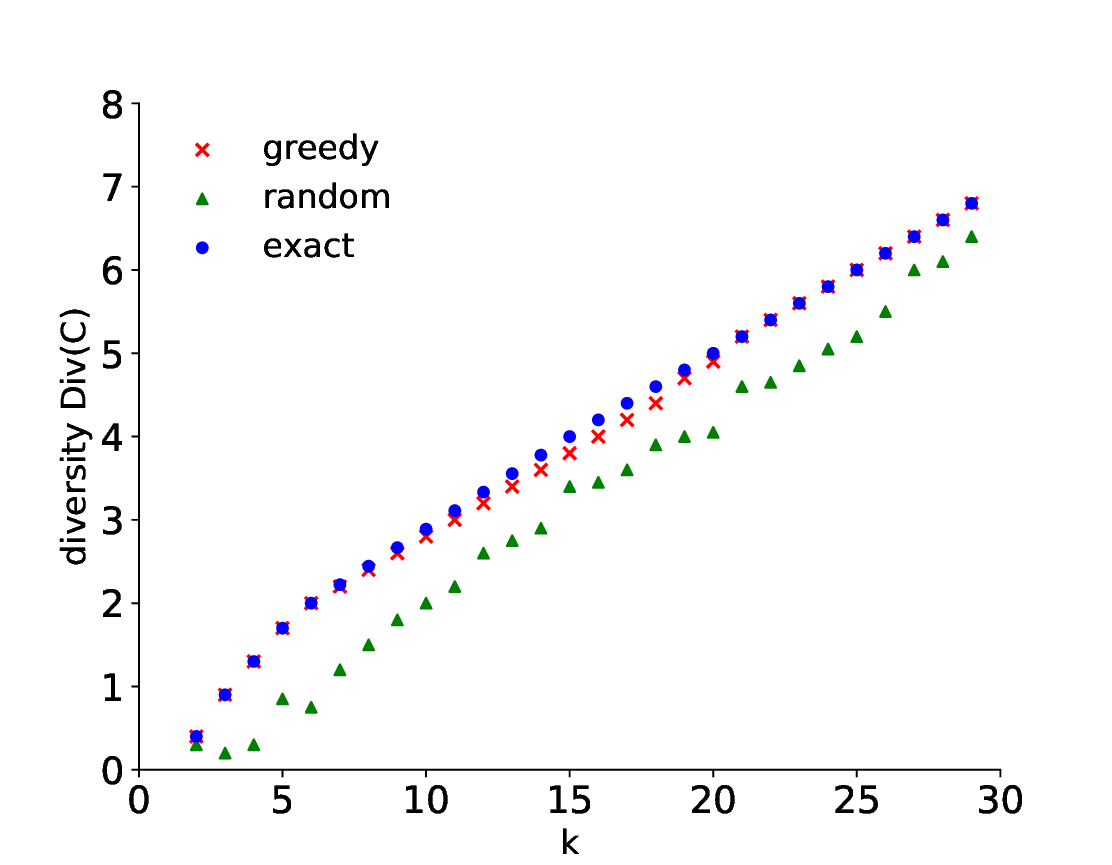}}
  \subfloat[Normal Distribution]{\includegraphics[scale=0.4]{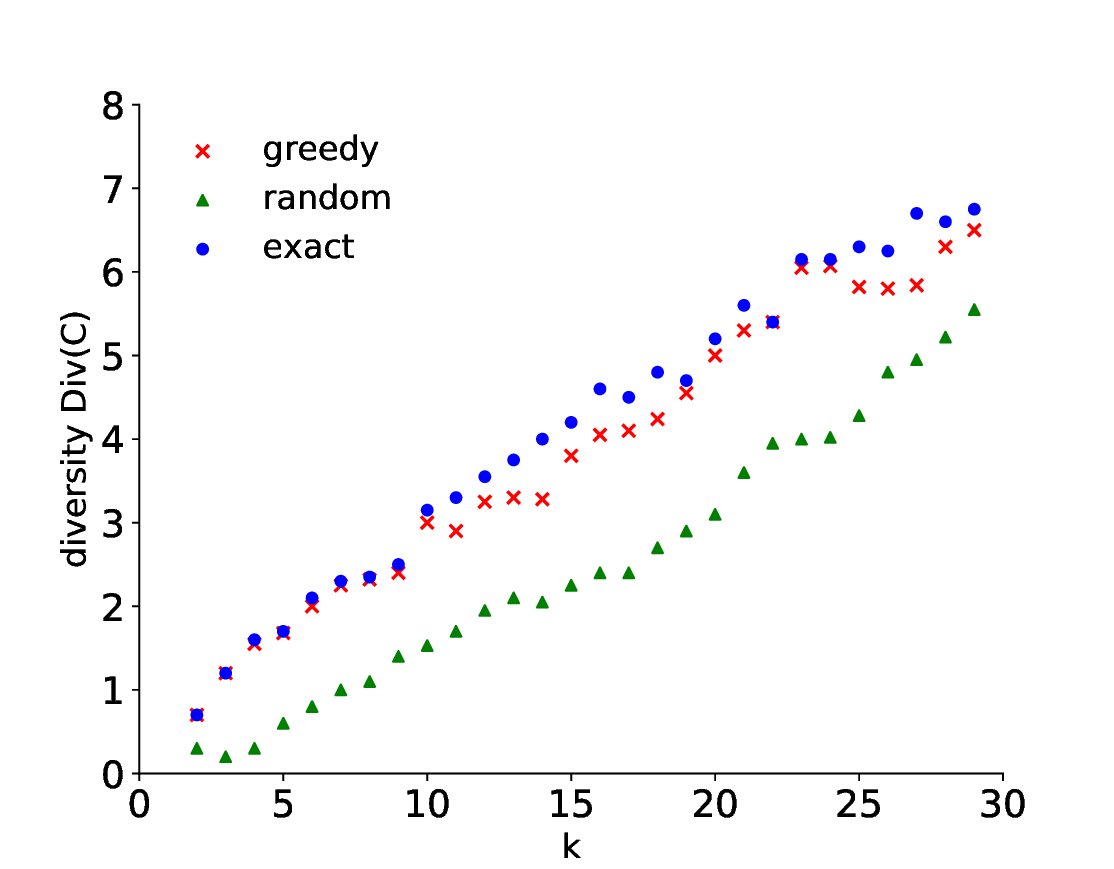}}
\caption{Effectiveness of Methods for S-model with Various Distributions}
\label{fig:4}
\end{figure*}

We first conducted evaluation on S-model. In particular, we compared the proposed greedy algorithm, namely \textit{greedy}, with two alternative methods as the baselines - (1) \textit{exact}: a brute-force algorithm, which computes the exact optimal solution; (2) \textit{random}: the workers are selected randomly. Due to the high computational cost for the exact algorithm, we only generate a small data set with 30 workers. Each pair of workers is assigned a similarity ranging from -1 to 0 (so ${Div(C)} > 0$), following two different distributions - Uniform and Normal.\\
\indent \textbf{Effectiveness}: We generated 100 such data sets, and reported their average performance in Fig.\ref{fig:4}. Note the x-axis denotes the budget number of workers to be enlisted, and y-axis indicates the diversity of the selected crowd. The two baselines are referred to as the `exact' point set and the `random' point set. The former denotes the optimal combination derived from all possible combination scenarios, showcasing the highest $Div$ value attainable. Conversely, the latter represents the $Div$ value resulting from the selection of a combination of workers at random.
\\
\indent It is straightforward to interpret our findings: from the experimental results, we can see that \textit{greedy} well approximates the performance of the \textit{exact}. This is consistent with our theoretical analysis that \textit{greedy} performs an approximation guarantee of 63\%, as shown in Section \ref{Approximation the S-model}. In addition, \textit{greedy} outperforms random for all two distributions. We also find that the diversity grows with the increasing number of $k$ for all two algorithms, which confirms the fact that large crowds tend to have high diversity. Another interesting finding is that, by comparing it with \textit{random}, the advantages of \textit{greedy} are more evident in Normal distributions than in Uniform distributions. This is because Normal distribution are skewed, thereby \textit{random} is very likely to select the values around the mean, which leads to low diversity. \\
\indent Since Algorithm 1 operates as a greedy algorithm, its applicability may be limited across various scenarios. To identify scenarios where the algorithm's performance is not good, we conducted tests using the MIN-SIM method. We set the number of candidate worker sets to 10 and $k$ to 6, generating 10,000 sets of random data. We then compared the results obtained from Algorithm 1 with the optimal solutions. Among these, we observed that the approximation rate for 20 sets of data fell below 80\% (i.e., $Div(C)/Div(C_{opt}) < 80\%$). Upon scrutinizing these results, we inferred that the algorithm's efficacy deteriorates when the similarity matrix contains a extreme small value and the corresponding $Sim$ value for the combination associated with this extreme small value is excessively large.\\
\indent This inference aligns with the algorithm's functioning. Given the initial selection of the combination with the minimum $Sim$ value, subsequent selections by the greedy algorithm are contingent upon this initial set of workers. Consequently, even if the greedy algorithm automatically selects new workers that maximize $Div$ value, the resulting $Div$ value may still be inferior to that of some other combinations due to the high $Sim$ values between the initially selected workers and others.\\
\indent In response to this observation, we also introduced an alternative initialization method called MIN-SUM in Section \ref{Approximation the S-model}. The MIN-SUM method aims to mitigate performance degradation stemming from extreme values.\\
\indent We conducted tests using both initialization schemes, and the results are presented in Table 4. In this table, `Methods' denotes the selection strategies, `Apr' represents the average approximation rate calculated by $Div(C)/Div(C_{opt})$, `Opt' indicates the rate of cases achieving the optimal solution, and `low' indicates the number of cases with an approximation rate lower than 80\%. We find that compared to the enumeration algorithm, the use of the MIN-SIM method led to more cases achieving results entirely consistent with the optimal solution. However, its overall approximation rate was lower. Conversely, when initializing with the MIN-SUM method, although the number of optimal solutions decreased, the overall approximation rate improved, and the number of bad cases becomes to 0.
\begin{table}[ht]\centering
\label{Experiments on two methods about S model}
\begin{tabular}{|c|c|c|c|}
\hline
Methods & Apr(\%) & Opt(\%) & Low \\ \hline
MIN-SIM & 97.72 & 25.3 & 20   \\\hline
MIN-SUM & 98.82 & 24.3 & 0  \\\hline
\end{tabular}
\caption{Efficiency of Methods for S-model with two Initial Selection Methods on Synthetic Data}
\end{table}
\subsubsection{Real data}
\label{Real data experiments on S-model}
\begin{figure}[h]
\centering
  \includegraphics[scale=0.4]{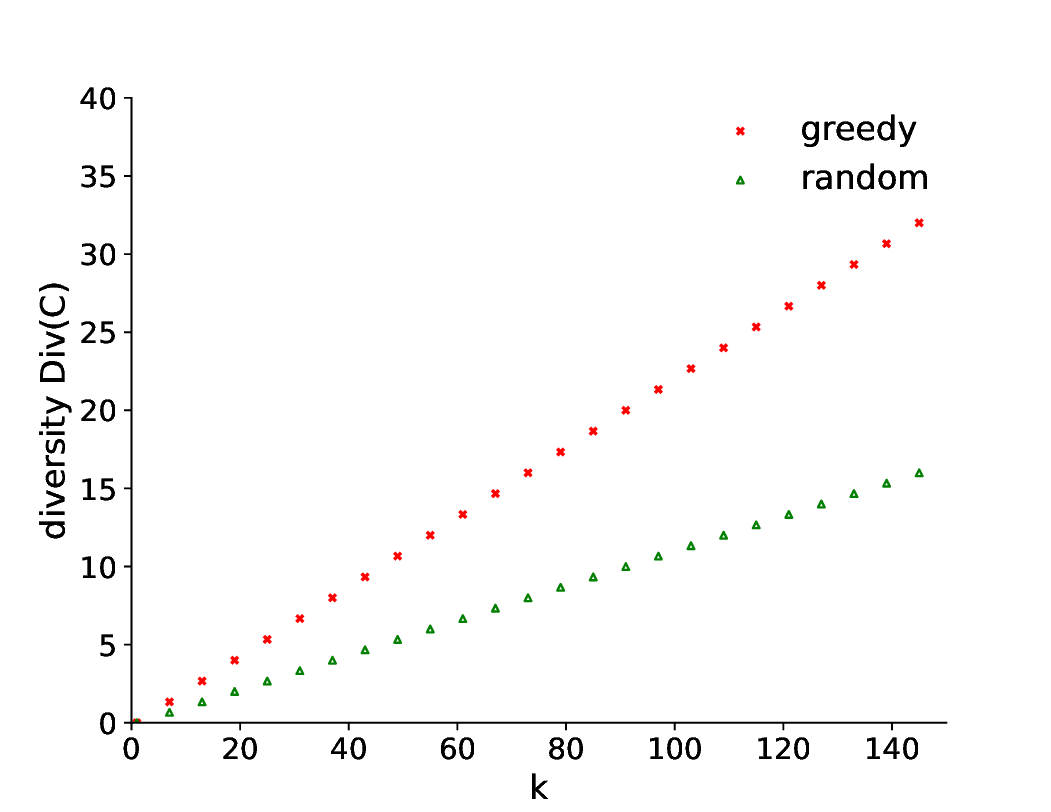}
\caption{Effectiveness of S-model on Foursquare (Real) data}
\label{fig:6}       
\end{figure}

\indent On the real data set, the exact algorithm cannot be performed due to its factorial time cost. So we only plotted the performance of \textit{random} and \textit{greedy}, as demonstrated in Fig.\ref{fig:6}, where there's only one baseline that is `random'. The result is basically consistent with the synthetic data.\\

\begin{figure*}[h]
\centering
  \subfloat[Uniform Distribution]{\includegraphics[scale=0.32]{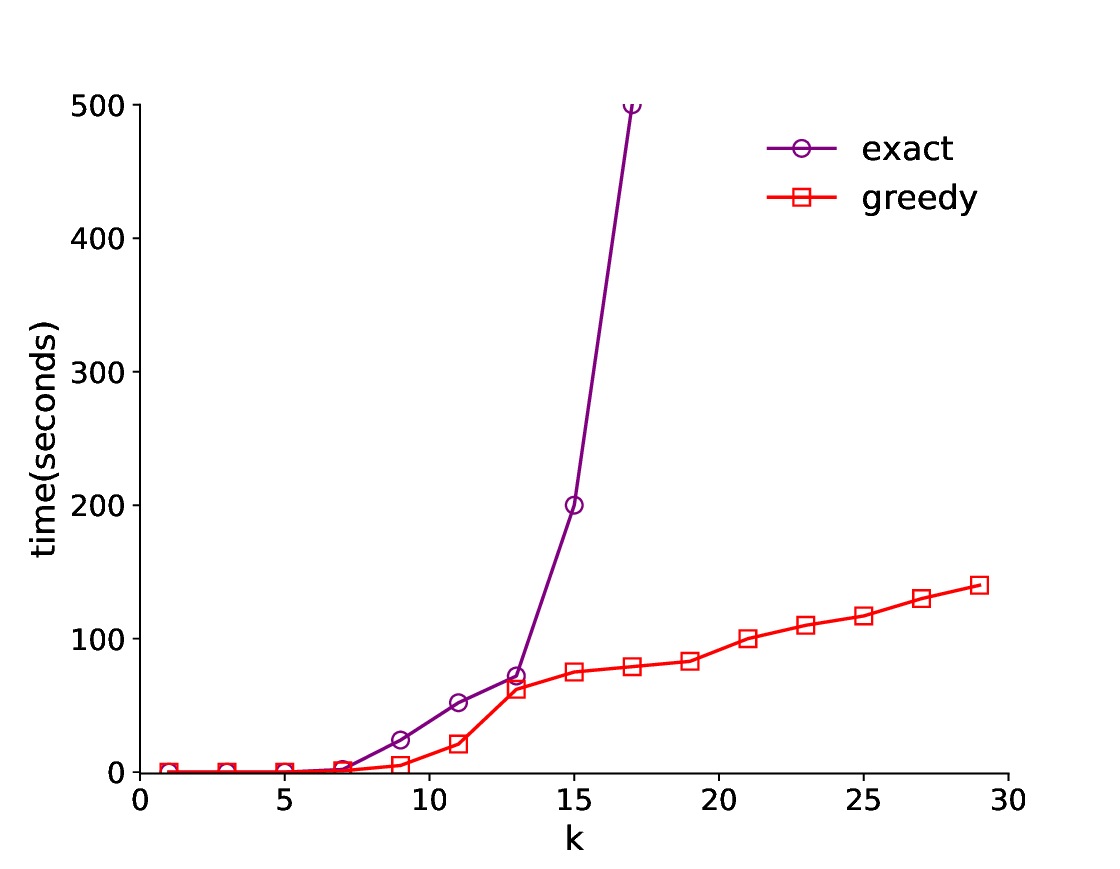}}
  \subfloat[Normal Distribution]{\includegraphics[scale=0.32]{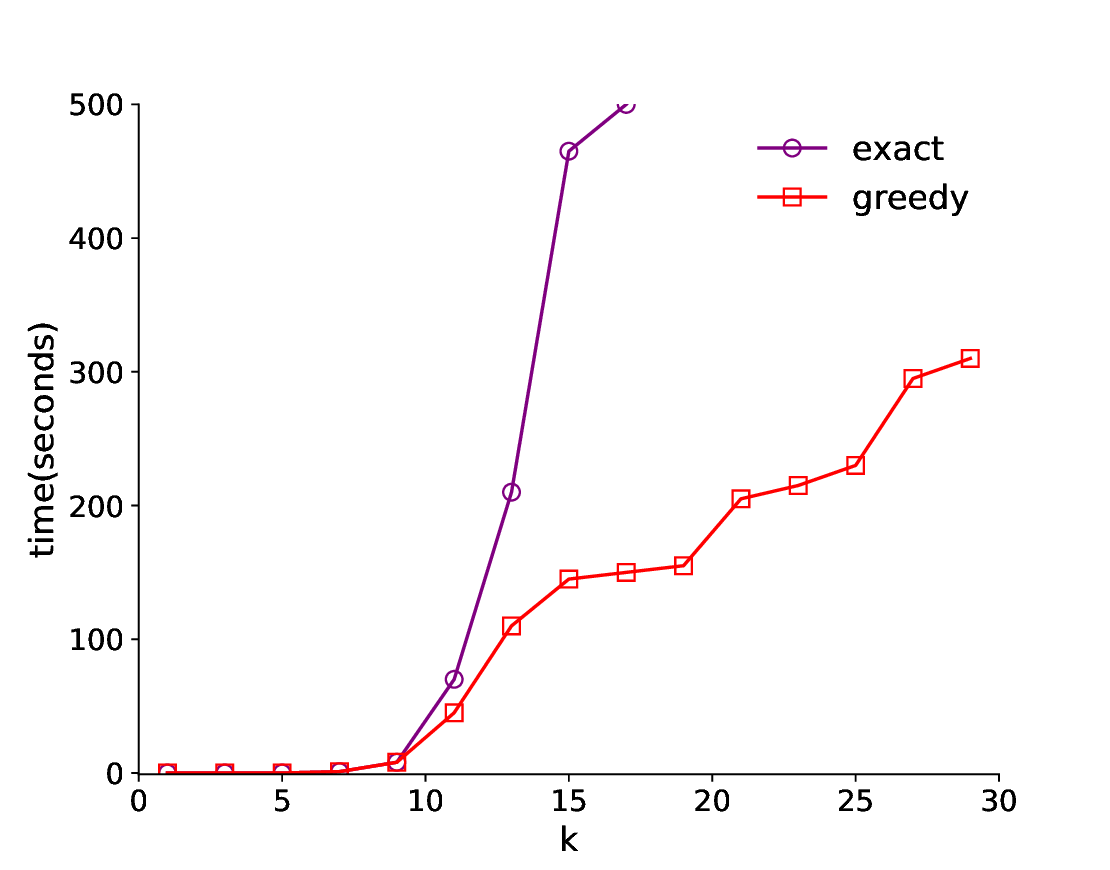}}
\caption{Efficiency of Methods for S-model with two Distributions}
\label{fig:5}
\end{figure*}

\indent \textbf{Efficiency}: In this subsection, we empirically examine the time-efficiency of the proposed algorithm for S-model. In particular, we compare the greedy algorithm (Algorithm 2) with the exact algorithm (Brute-force enumeration). As shown in Fig.\ref{fig:5}, the exact algorithm (denoted by \textit{exact} and it's the baseline of this figure) entails exponential computation time, and the greedy algorithm (\textit{greedy}) is much more effective than $exact$. Please note that we stop $exact$ after running it over 500 seconds.\\
\begin{figure*}[h]
\centering
  \subfloat[Case 1: 100 candidates]{\includegraphics[scale=0.28]{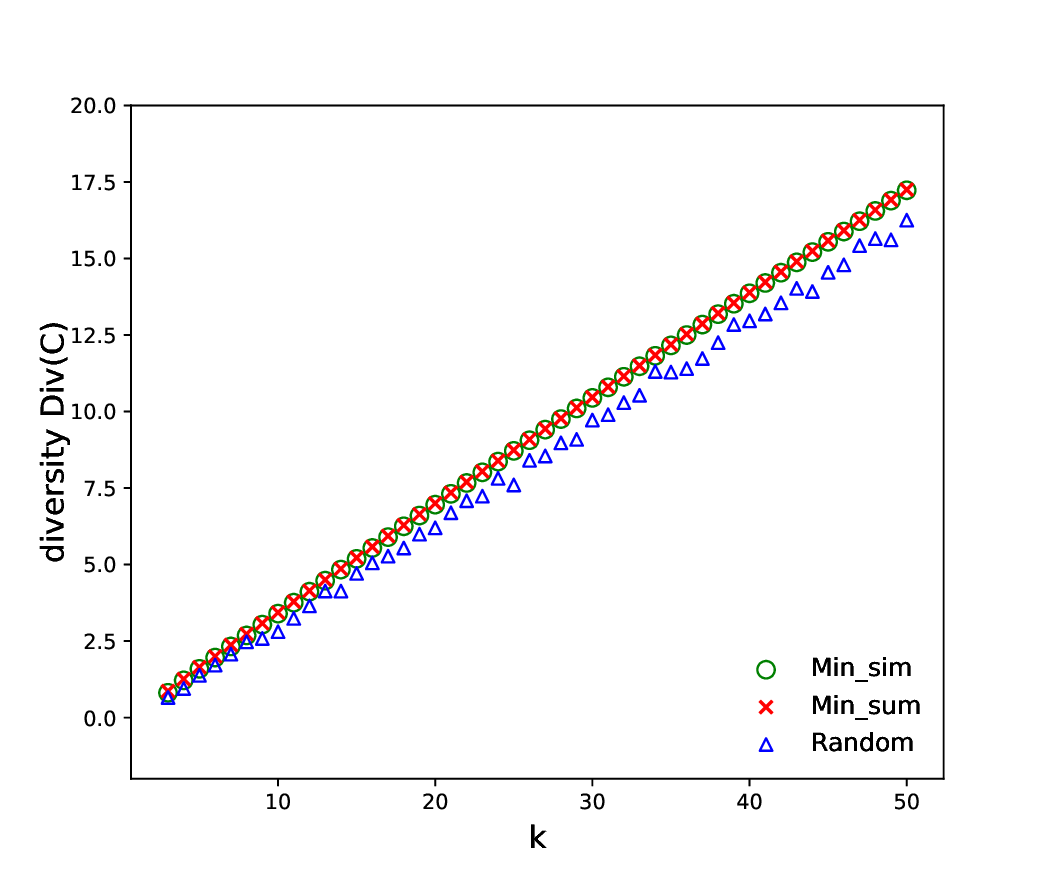}}
  \subfloat[Case 2: 200 candidates]{\includegraphics[scale=0.28]{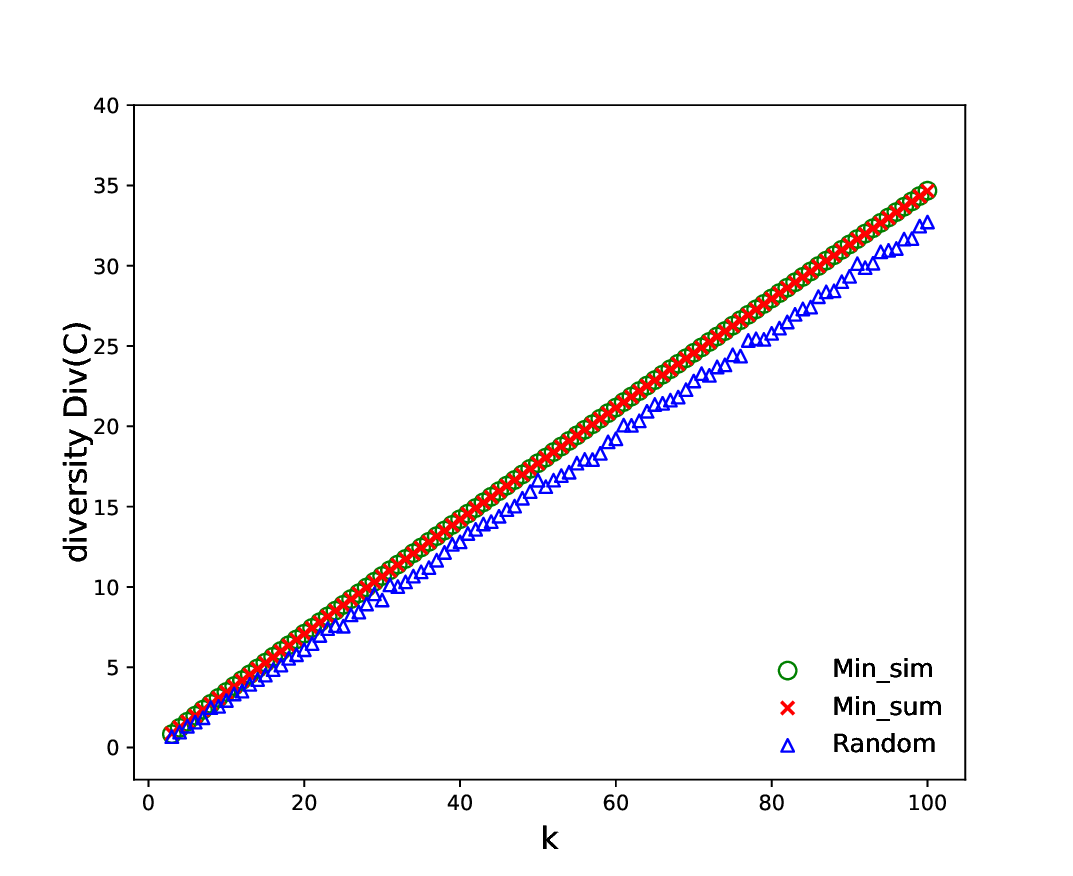}}
  \subfloat[Case 3: 400 candidates]{\includegraphics[scale=0.28]{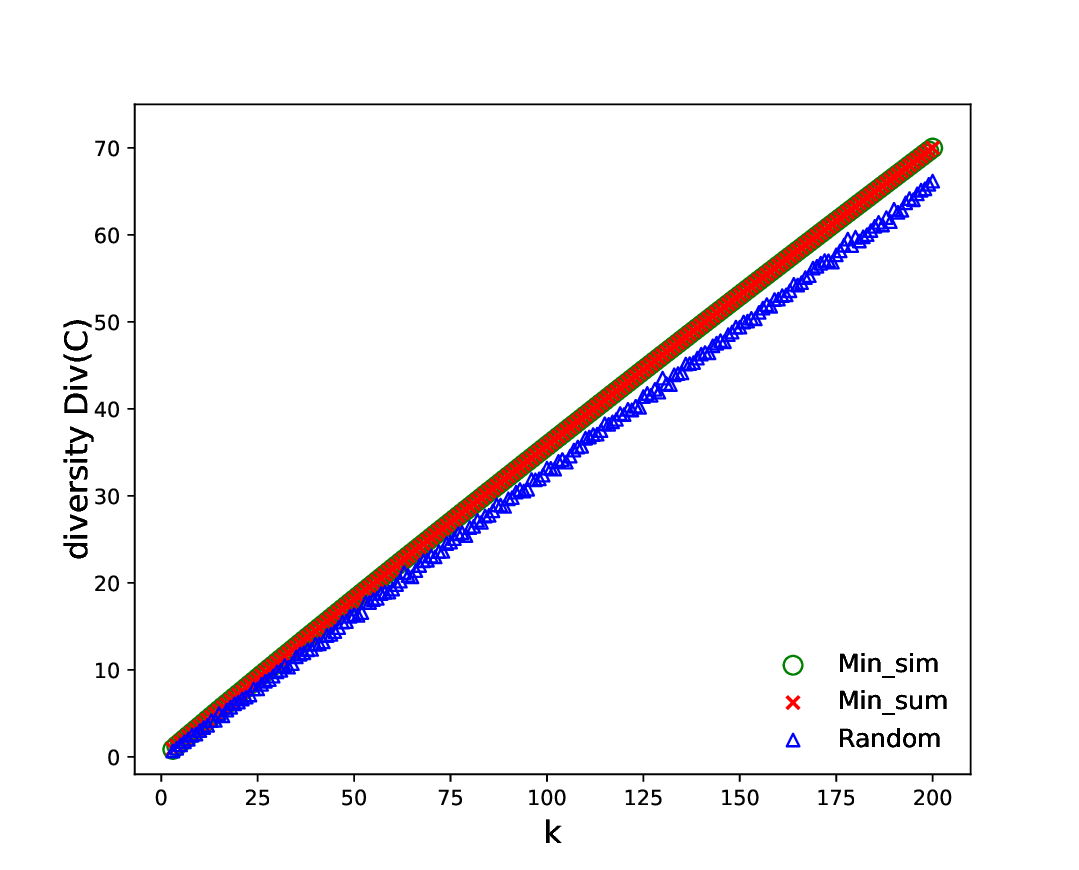}}
\caption{Efficiency of Methods for S-model with two Initial Selection Methods under three cases in Real Data}
\label{fig:6.5}
\end{figure*}
\indent Furthermore, we conducted experimental tests to investigate the effects of two different initialization selection methods on real data. Specifically, we utilized the Music \& Mental Health Survey Results dataset \cite{Ref53}, randomly selecting 100, 200 and 400 investigators as the candidate sets, respectively. We then employed the Jaccard distance, as described in Section \ref{Pairwise profile-based diversity}, to calculate the $Sim$ value between each investigator based on their 30 features in the dataset. Subsequently, we utilized the S-model to select varying $k$ of investigators and calculated their corresponding $Div$ values. The results are shown in Fig.\ref{fig:6.5}. We can see that for all of the three cases, as the parameter $k$ increases incrementally, the outcomes generated by the greedy algorithms under `MIN\_SIM' and `MIN\_SUM' consistently surpass those obtained by the random algorithm.
\subsection{Experiments on T-model}
\label{Experiments on T-model}
\subsubsection{Synthetic Data}
\label{Experiments on synthetic data}
\begin{figure*}
\centering
  \subfloat[Uniform \emph{k}=10,$\theta_1$=3,$\theta_0$=3]{\includegraphics[scale=0.26]{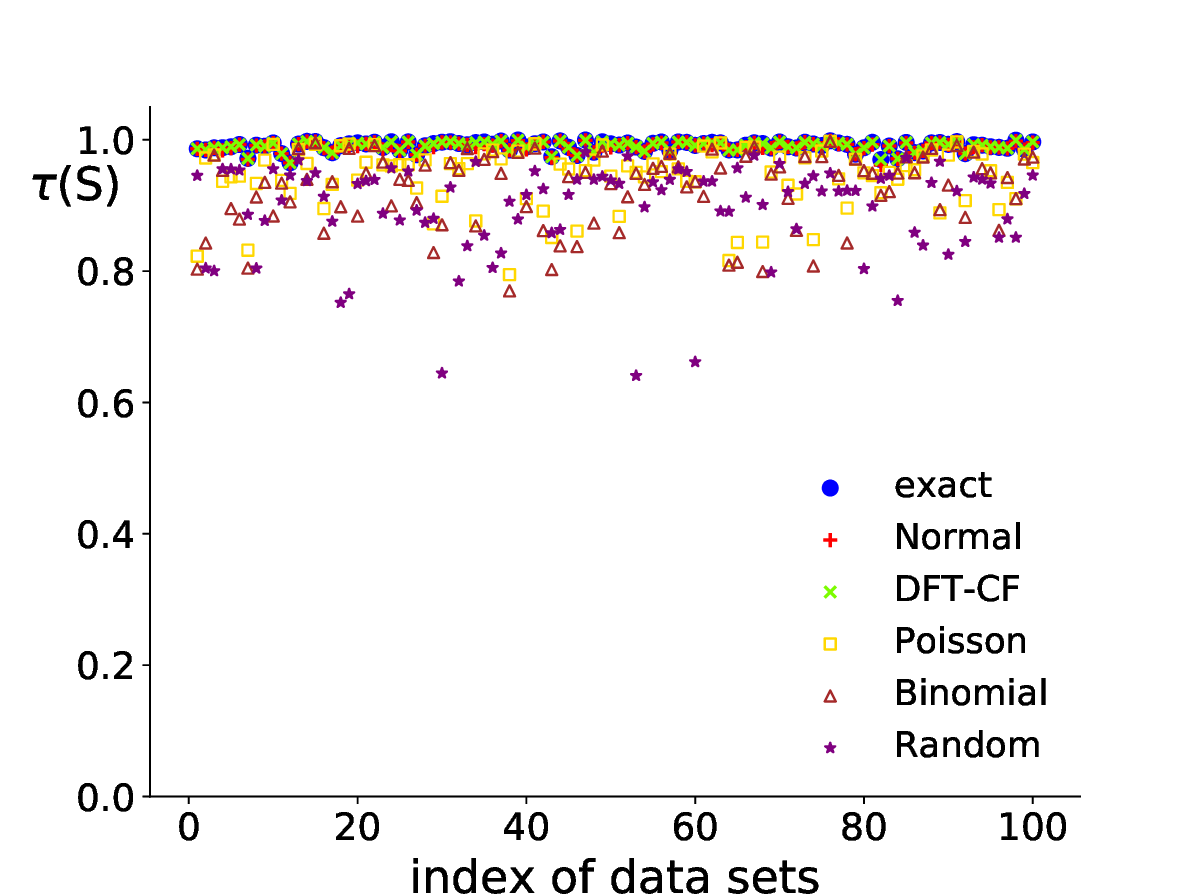}}
  \subfloat[Uniform \emph{k}=15,$\theta_1$=5,$\theta_0$=5]{\includegraphics[scale=0.26]{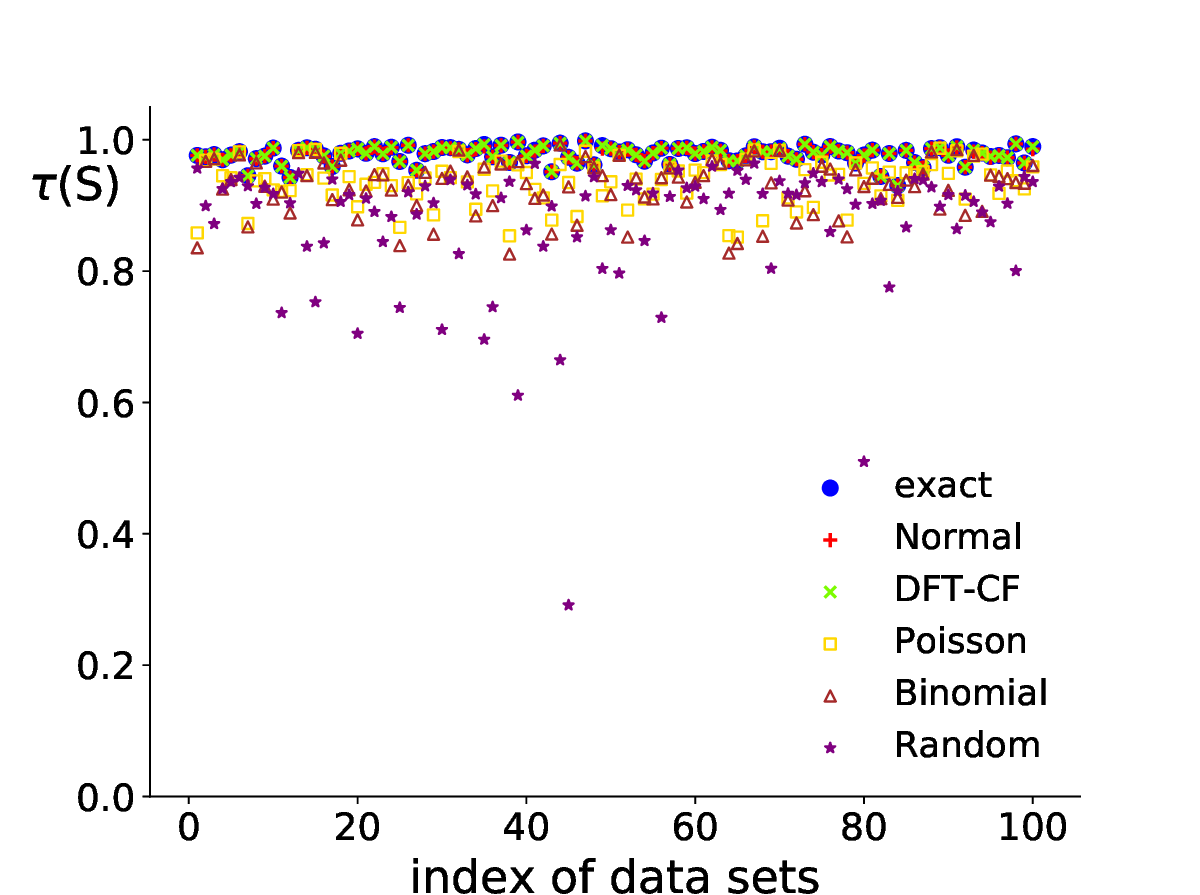}}
  \subfloat[Uniform \emph{k}=20,$\theta_1$=6,$\theta_0$=6]{\includegraphics[scale=0.26]{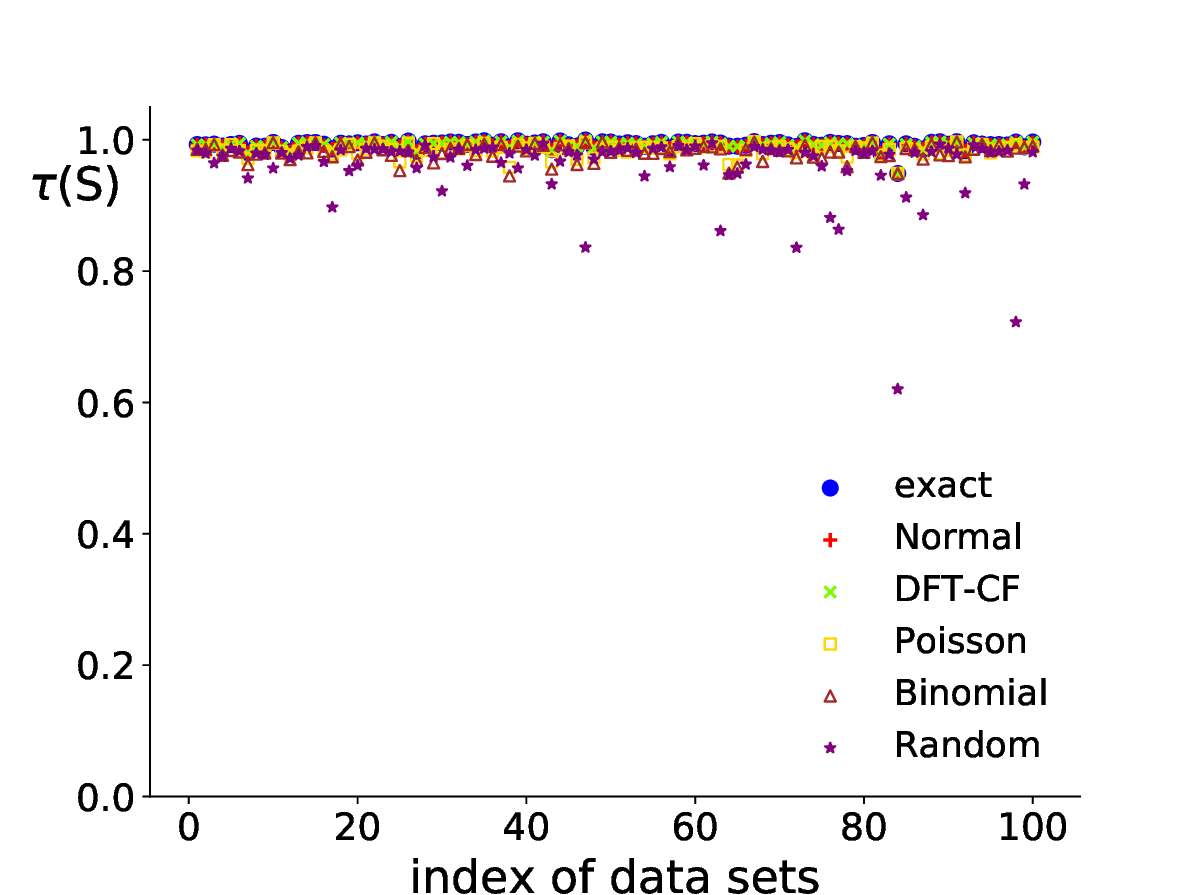}}\\
  \subfloat[Normal \emph{k}=10,$\theta_1$=3,$\theta_0$=3]{\includegraphics[scale=0.26]{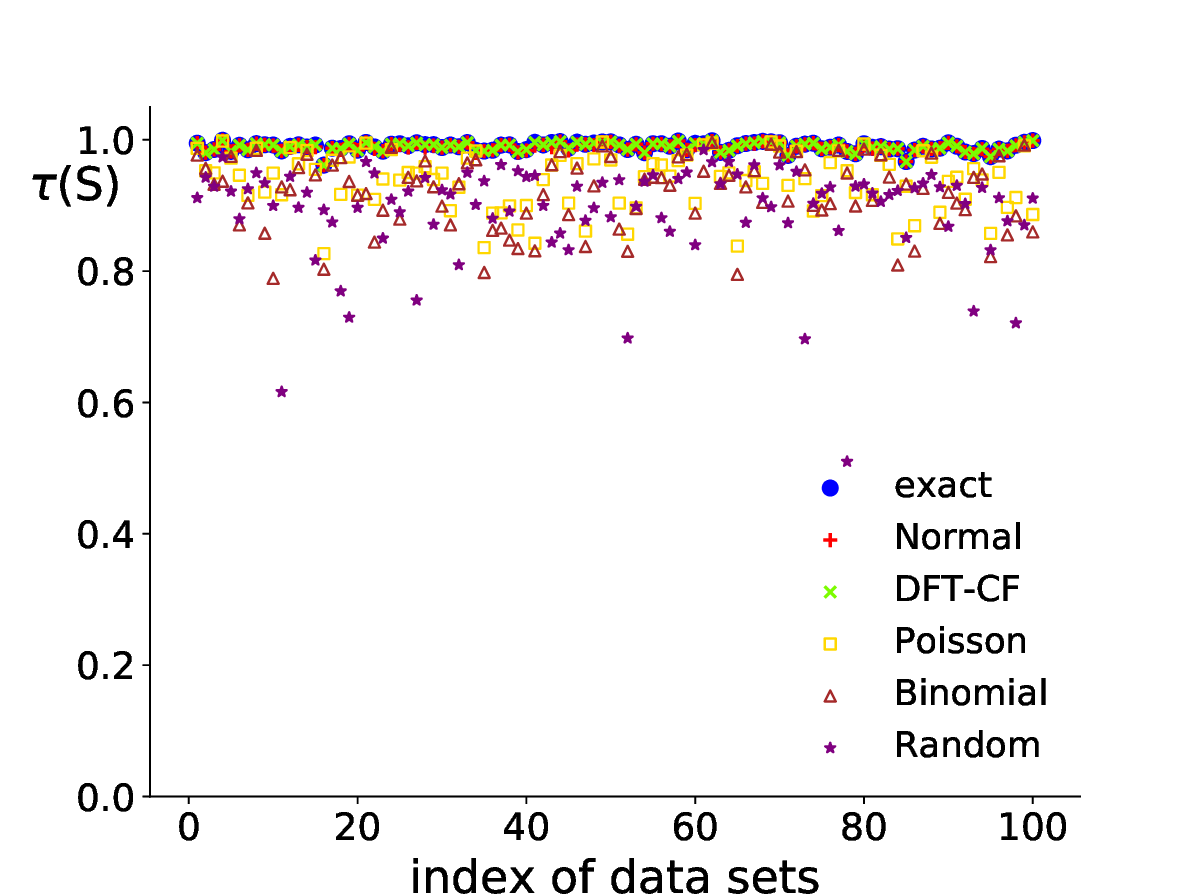}}
  \subfloat[Normal \emph{k}=15,$\theta_1$=5,$\theta_0$=5]{\includegraphics[scale=0.26]{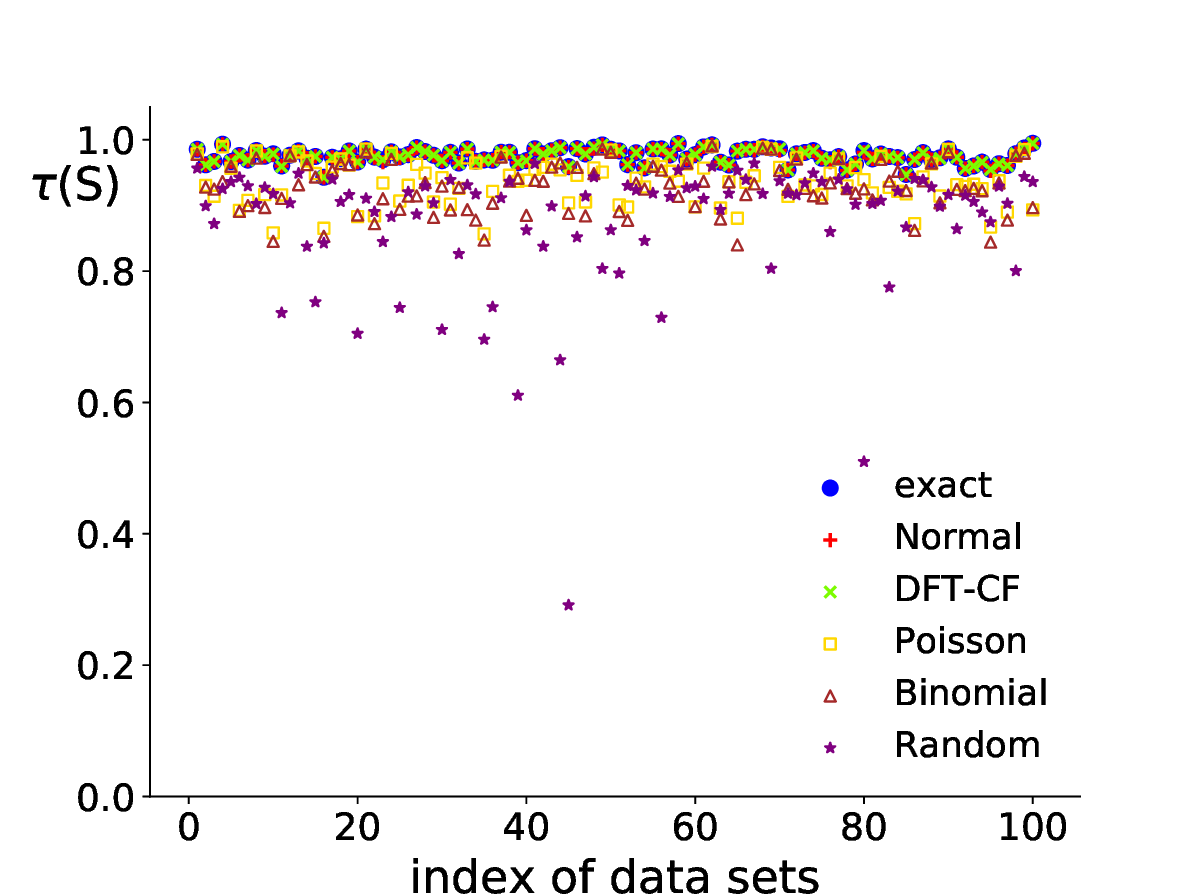}}
  \subfloat[Normal \emph{k}=20,$\theta_1$=6,$\theta_0$=6]{\includegraphics[scale=0.26]{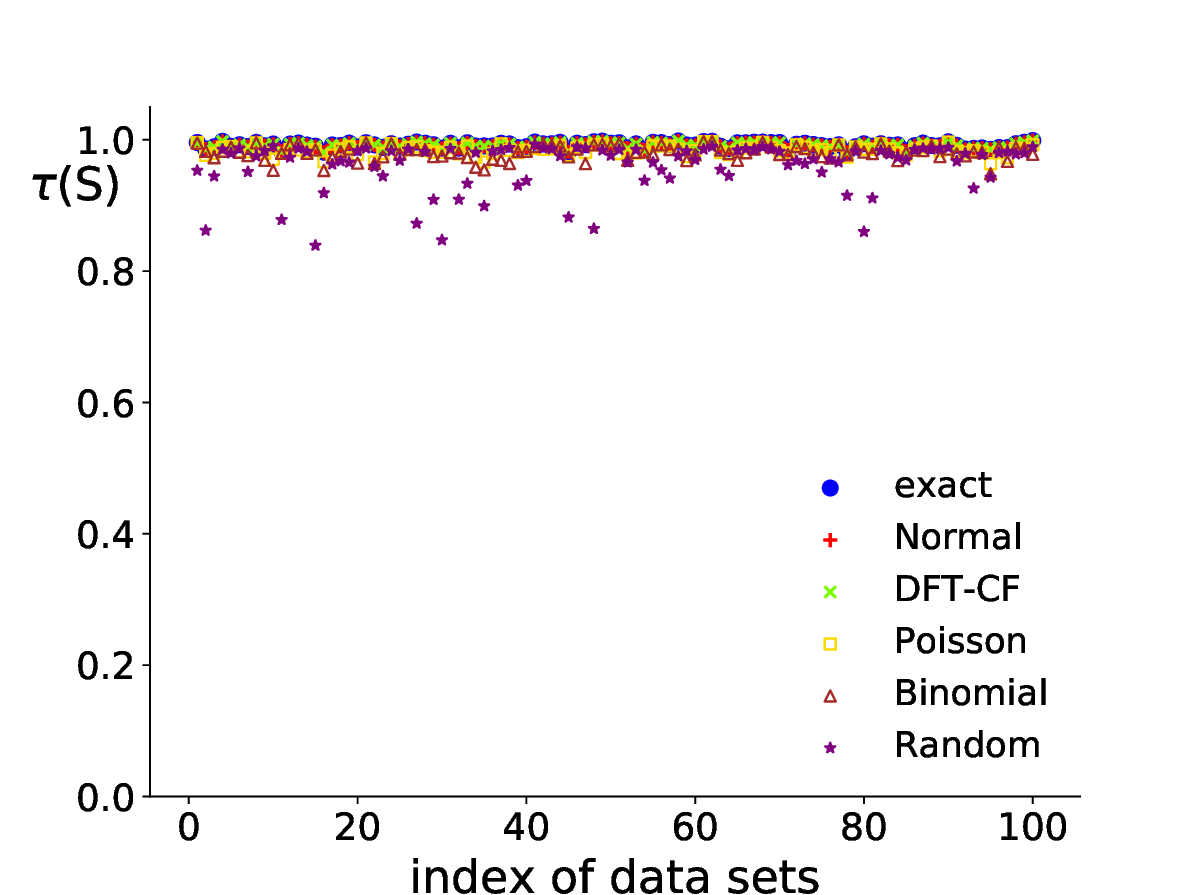}}\\
  \subfloat[Beta \emph{k}=10,$\theta_1$=3,$\theta_0$=3]{\includegraphics[scale=0.26]{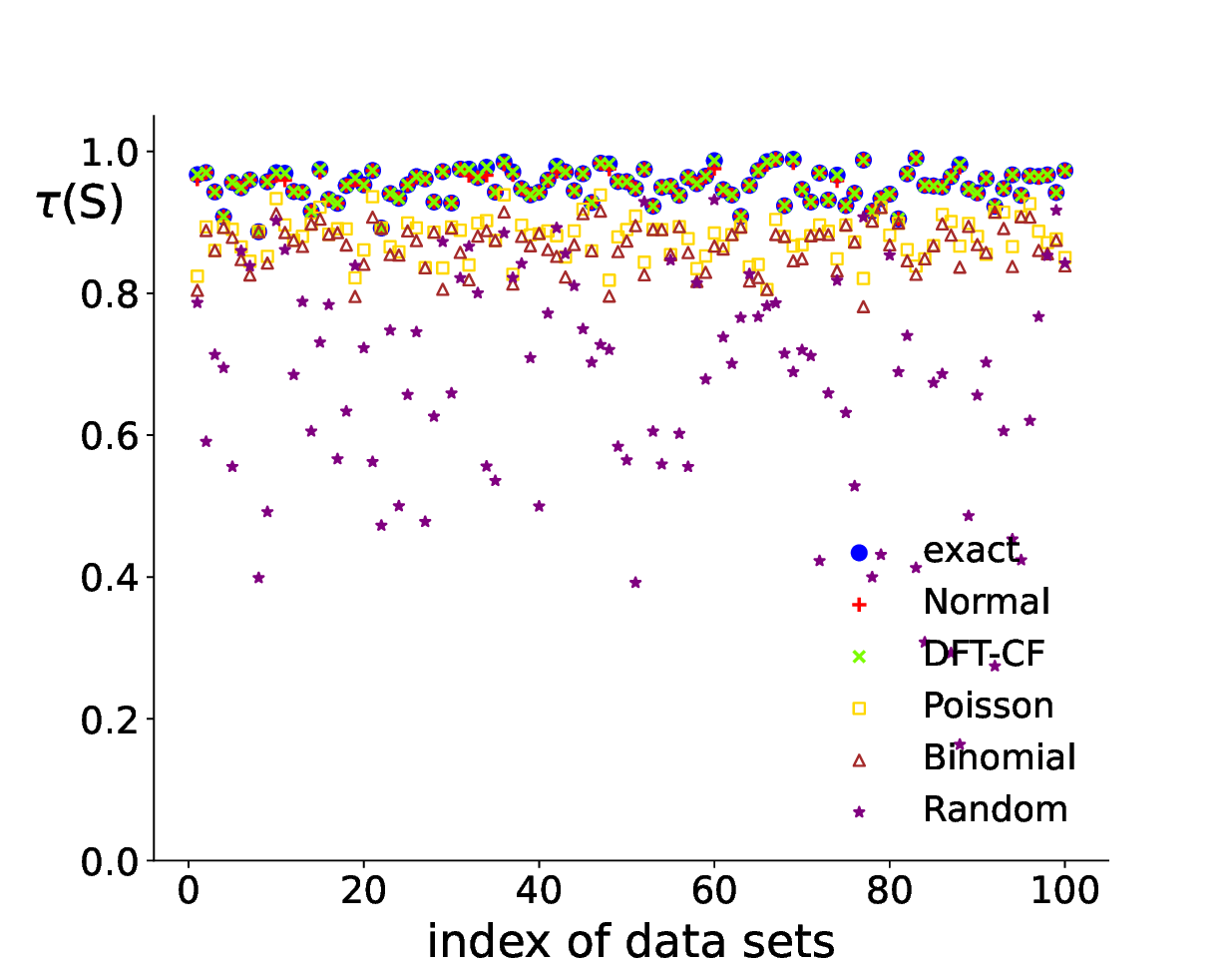}}
  \subfloat[Beta \emph{k}=15,$\theta_1$=5,$\theta_0$=5]{\includegraphics[scale=0.26]{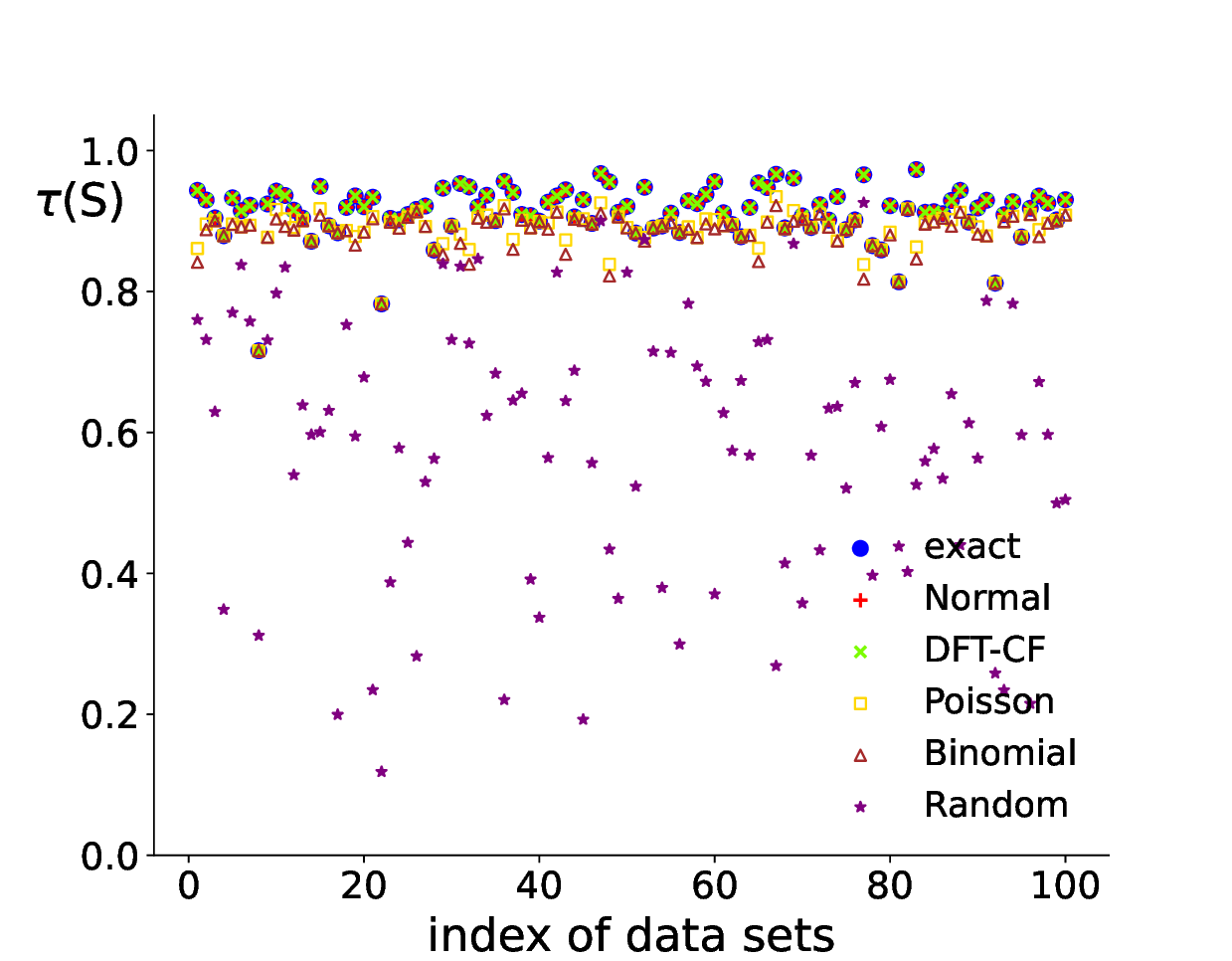}}
  \subfloat[Beta \emph{k}=20,$\theta_1$=6,$\theta_0$=6]{\includegraphics[scale=0.26]{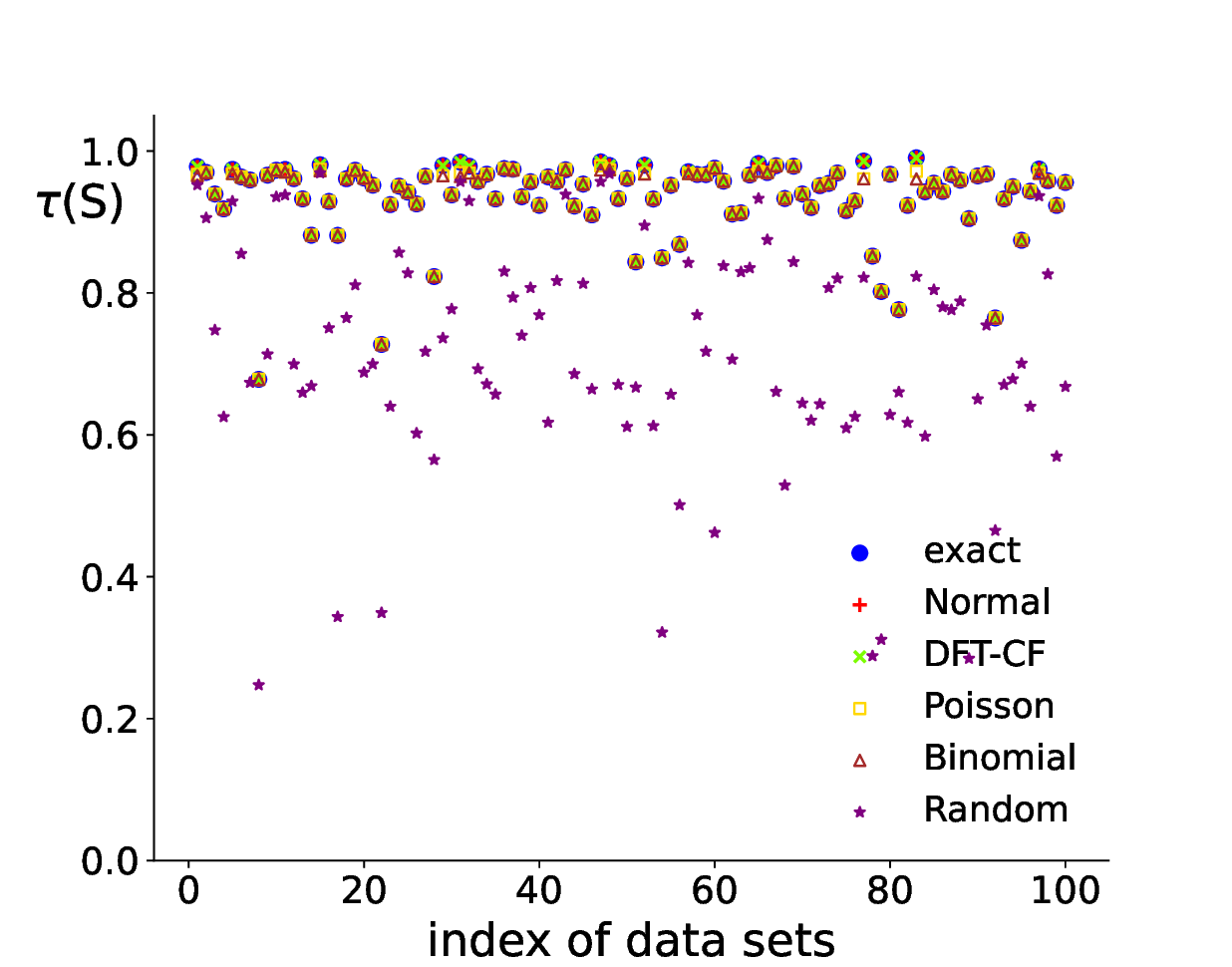}}
\caption{Effectiveness of Methods with Poisson, Binomial, Normal and DFT-CF Approximations}
\label{fig:7}
\end{figure*}
\begin{figure*}
\centering
  \subfloat[Uniform Distribution]{\includegraphics[scale=0.35]{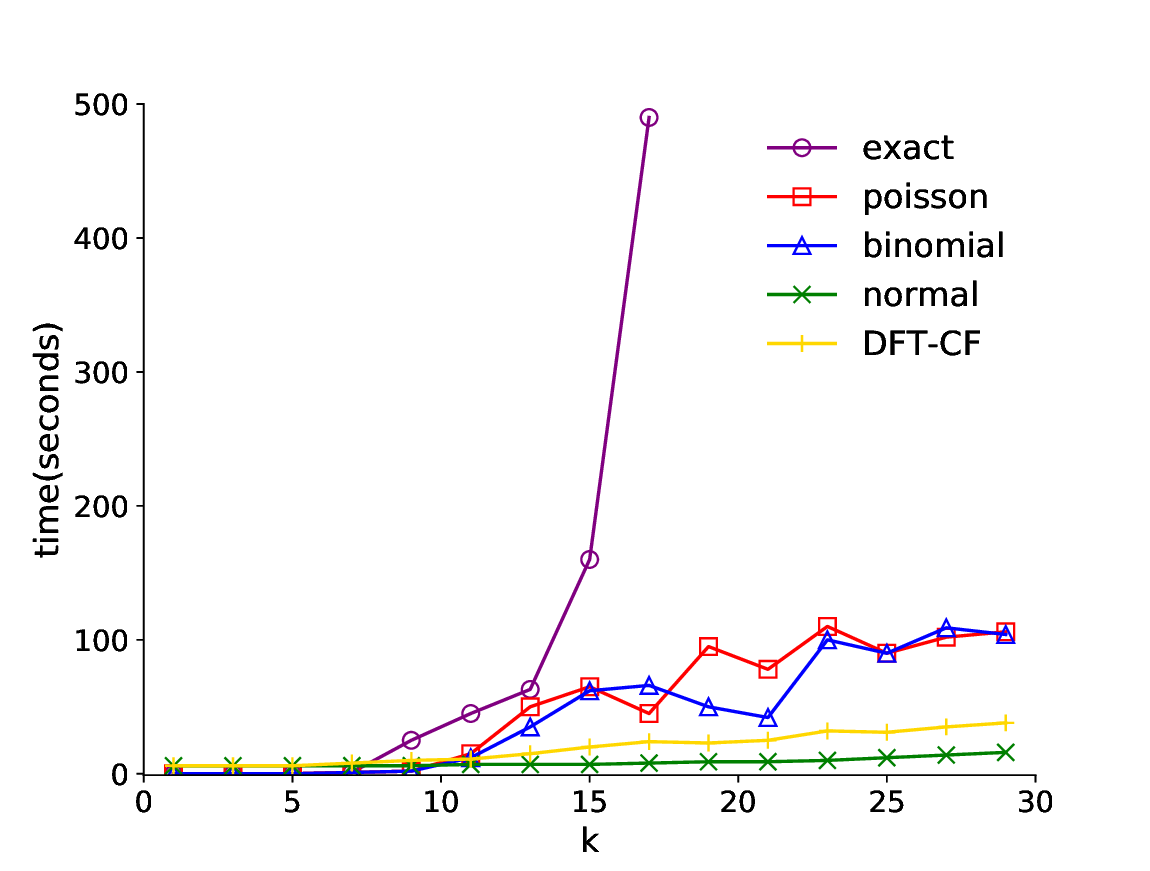}}
  \subfloat[Normal Distribution]{\includegraphics[scale=0.35]{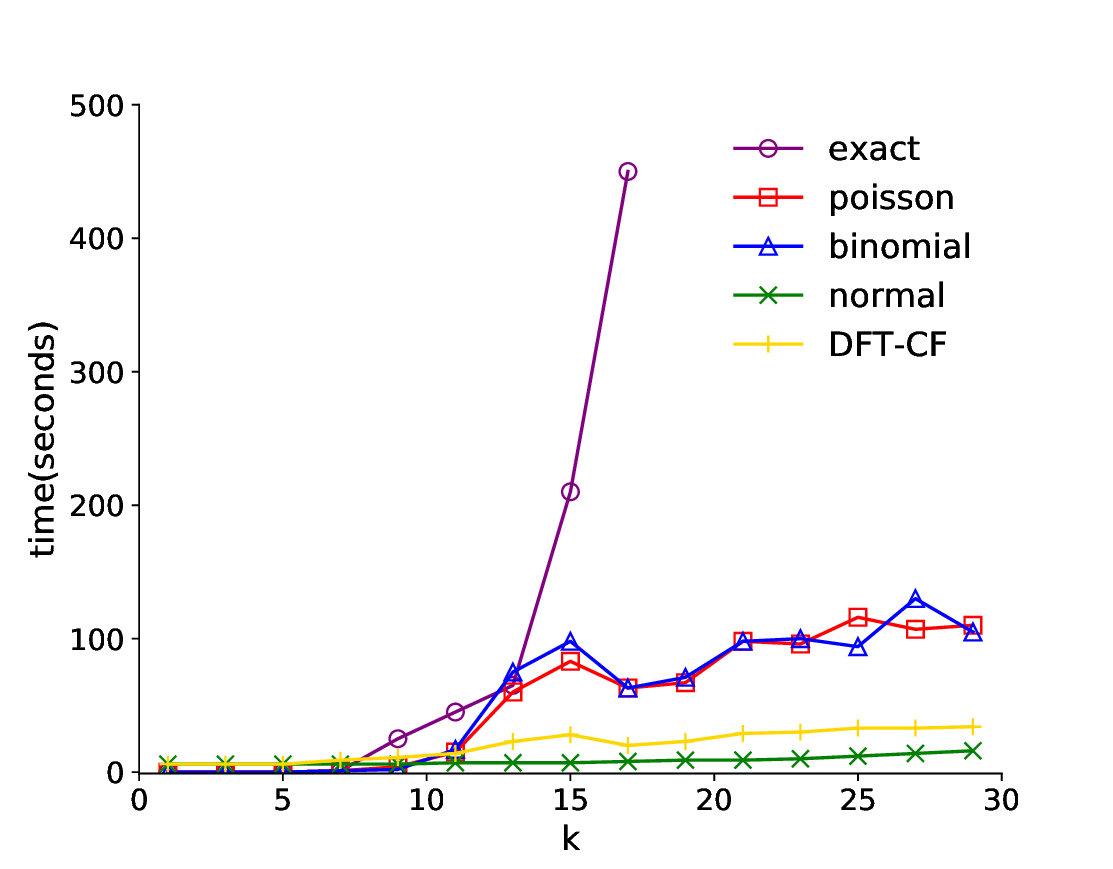}}
\caption{Efficiency of Methods for T-model with Various Distributions}
\label{fig:8}
\end{figure*}
In this subsection, we demonstrate a series of experimental results on synthetic data. To simulate individual opinions without bias, in this section we produced synthetic datasets following three different distributions: the first two are uniform distribution and normal distribution, both with varying mean and variance, and the third one is beta distribution with $\alpha =1$ and $\beta=2$. The characteristics of K-Best selection are investigated with both Poisson approximation and Binomial approximation and the characteristics of Simulated Annealing algorithms are investigated with both normal approximation and exact DFT-CF method. Then we evaluate the efficiency and effectiveness of both methods.\\
\indent The synthetic dataset is generated as follows: we generated 100 data sets, each including 30 candidate workers. The number of candidate workers is small because we want to use a brute-force algorithm to traverse the searching space, and find the absolute optimal solution. Then, we can evaluate how far the proposed approximation algorithm is from this optimum. The setting of parameters is: $k = 10,\theta_1 = 3,\theta_0 = 3$, $k = 15,\theta_1 = 5, \theta_0 = 5$ and $k = 20,\theta_1 = 6,\theta_0 = 6$.\\
\indent The results of effectiveness are reported in Fig.\ref{fig:7}. In each subfigure of Fig.\ref{fig:7}, x-axis indicates the index of the 100 data sets, and y-axis denotes the value of $\tau(S)$, which is the function we try to maximize. The methods with Poisson and Binomial approximations are named `Poisson' and `Binomial', and the methods with normal approximation and exact DFT-CF are named `Normal' and `DFT-CF', respectively. To better illustrate the advantage of the proposed methods, we also compare them with two baseline methods, which randomly select workers and select the optimal worker combination by enumerating all the possibilities respectively (denoted by `Random' and `exact', respectively). From the experimental results, we can see that the performance of `Random' can be arbitrarily bad, while the `Poisson' and `Binomial' have similar performance and the `Normal' and `DFT-CF' have better performance, and they all well approximate the optimum. In addition, we present the comparison of efficiency in Fig.\ref{fig:8}. One can see that the approximation techniques are much more efficient than computing the exact solutions. Moreover, we observe that `Poisson' and `Binomial' have similar performance in terms of efficiency, the method `Normal' performs the best, and `DFT-CF' ranks the second.
\subsubsection{Real Data}
\label{Experiments on real data}
\begin{figure*}
\centering
  \subfloat[$|N|$=10000,$\theta_1$=1,$\theta_0$=2\emph{k}/3]{\includegraphics[scale=0.28]{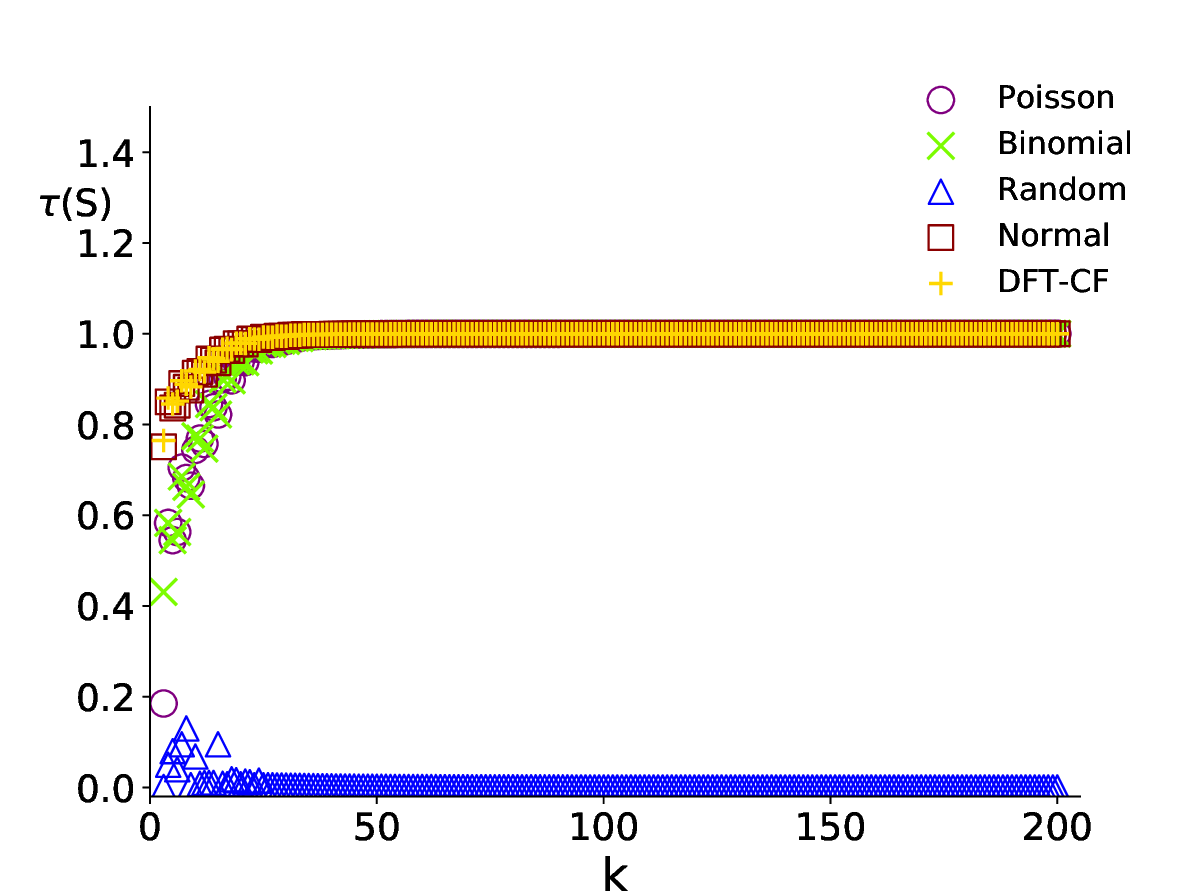}}
  \subfloat[$|N|$=10000,$\theta_1$=\emph{k}/3,$\theta_0$=\emph{k}/3]{\includegraphics[scale=0.28]{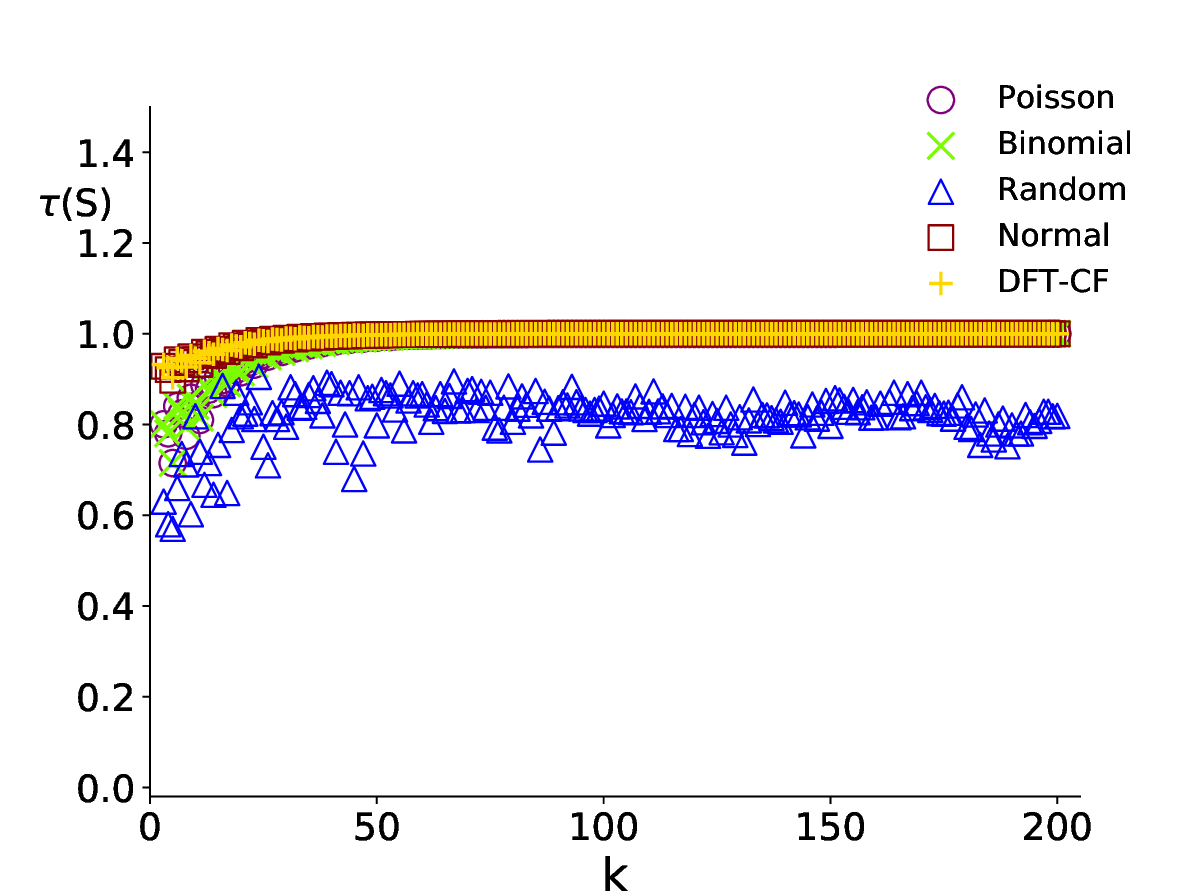}}
  \subfloat[$|N|$=10000,$\theta_1$=2\emph{k}/3,$\theta_0$=1]{\includegraphics[scale=0.28]{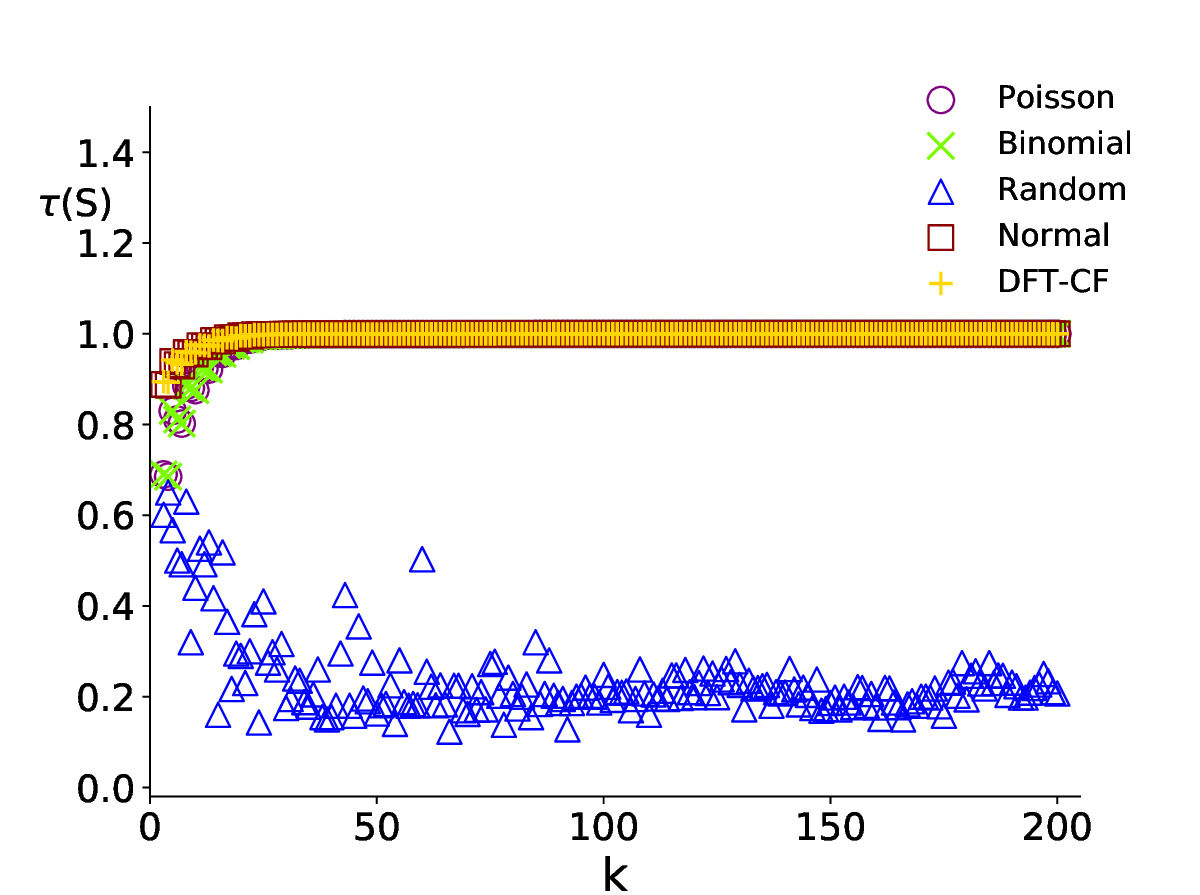}}\\
  \subfloat[$|N|$=10000,$\theta_1$=1,$\theta_0$=4\emph{k}/5]{\includegraphics[scale=0.28]{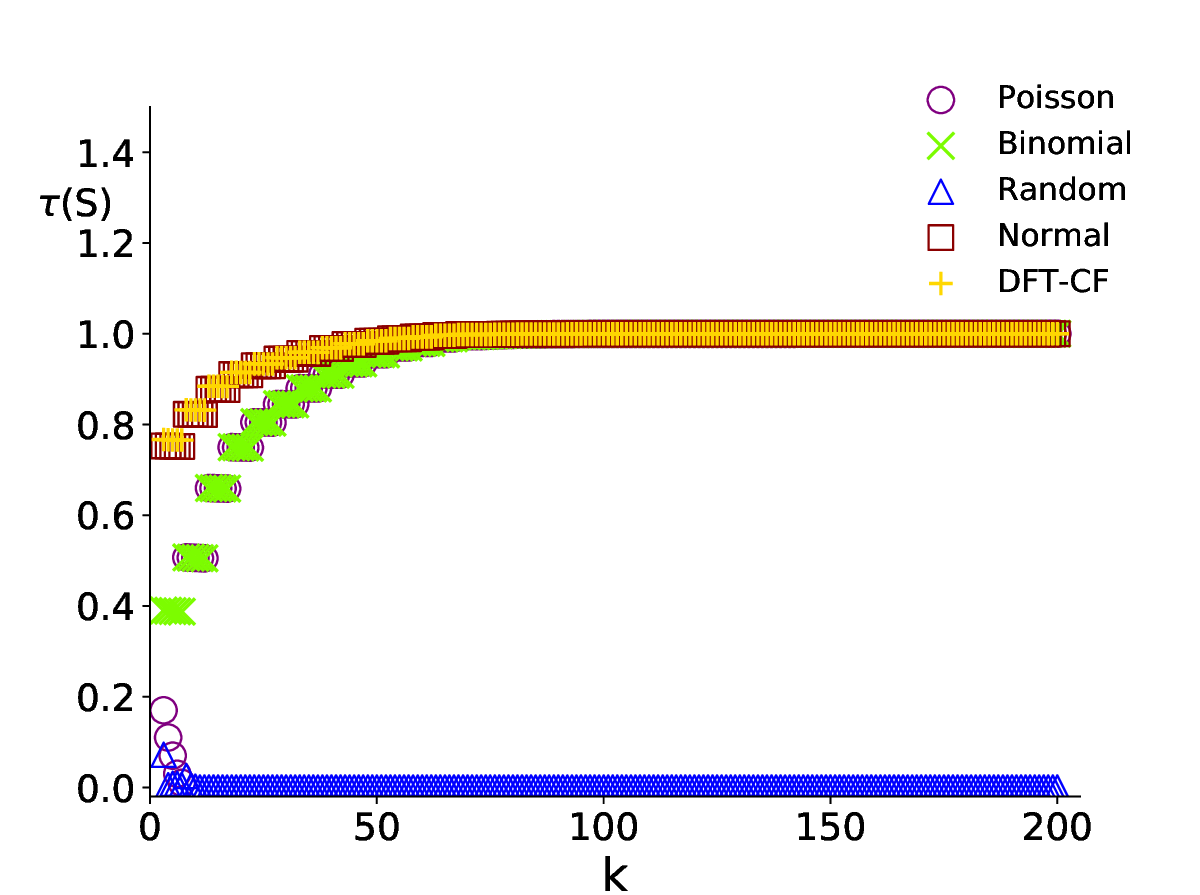}}
  \subfloat[$|N|$=10000,$\theta_1$=\emph{k}/5,$\theta_0$=3\emph{k}/5]{\includegraphics[scale=0.28]{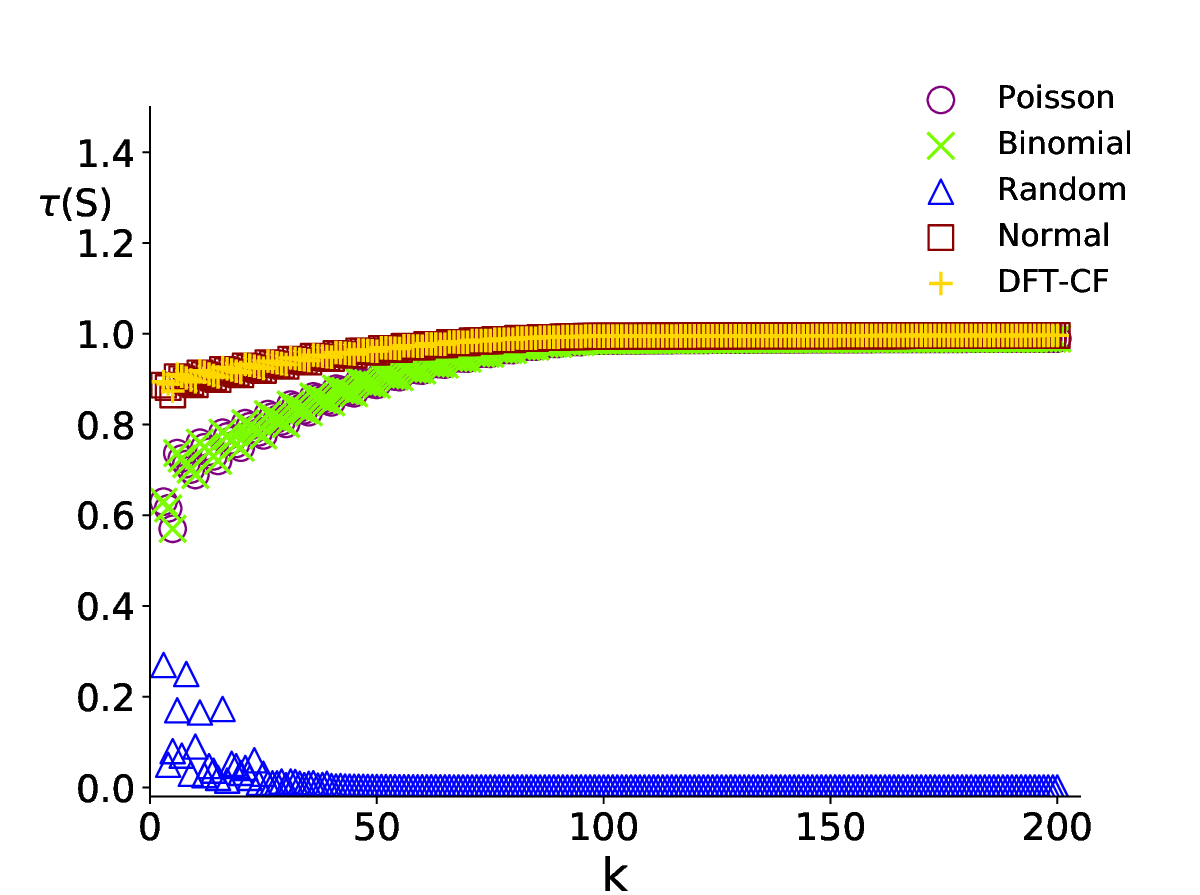}}
  \subfloat[$|N|$=10000,$\theta_1$=2\emph{k}/5,$\theta_0$=2\emph{k}/5]{\includegraphics[scale=0.28]{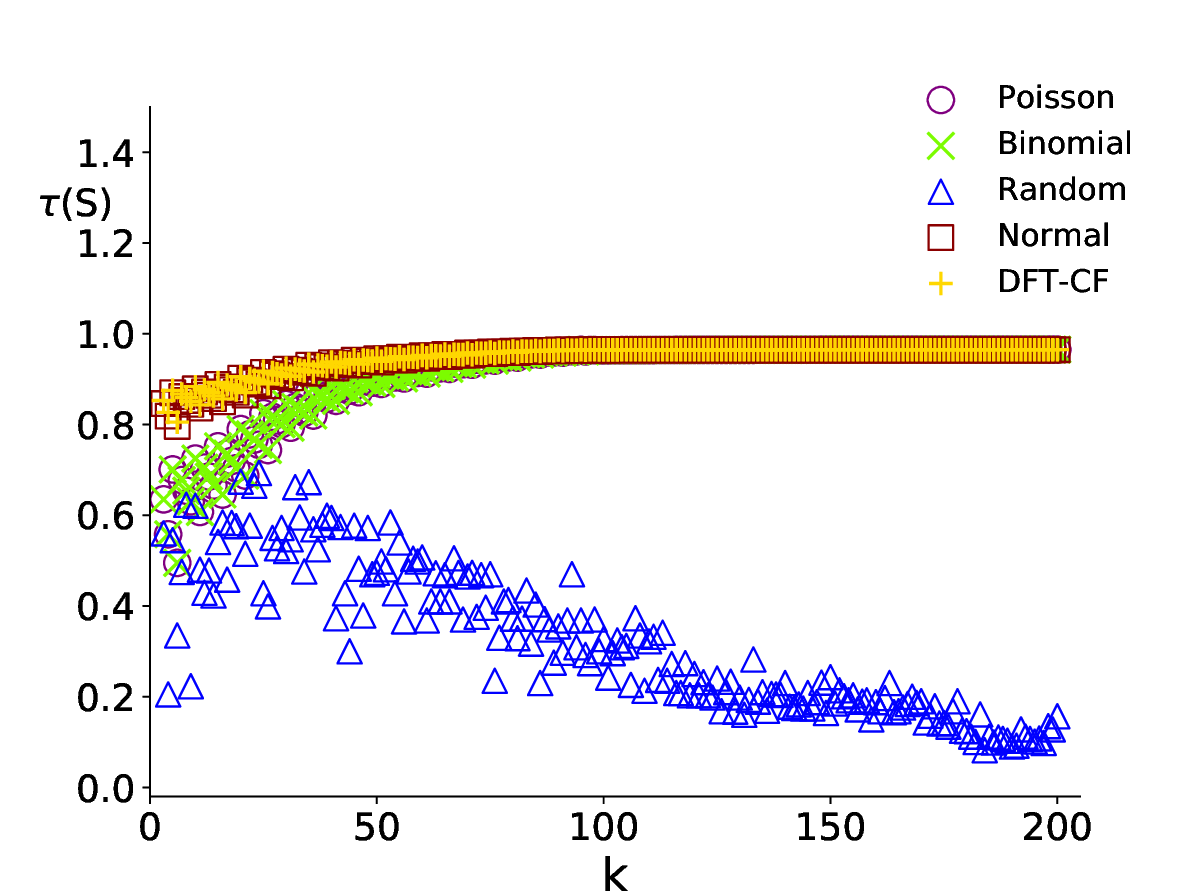}}\\
  \subfloat[$|N|$=10000,$\theta_1$=3\emph{k}/5,$\theta_0$=\emph{k}/5]{\includegraphics[scale=0.28]{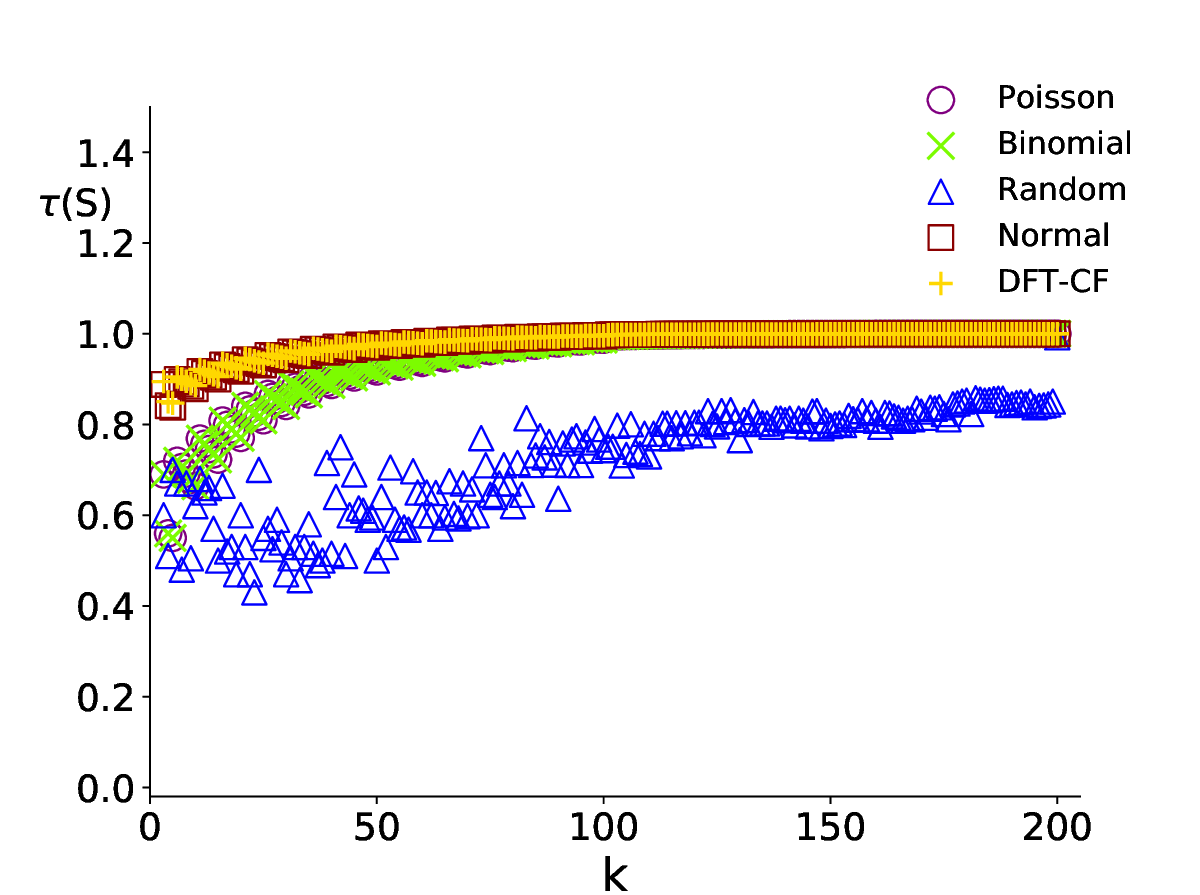}}
  \subfloat[$|N|$=10000,$\theta_1$=4\emph{k}/5,$\theta_0$=3\emph{k}/5]{\includegraphics[scale=0.28]{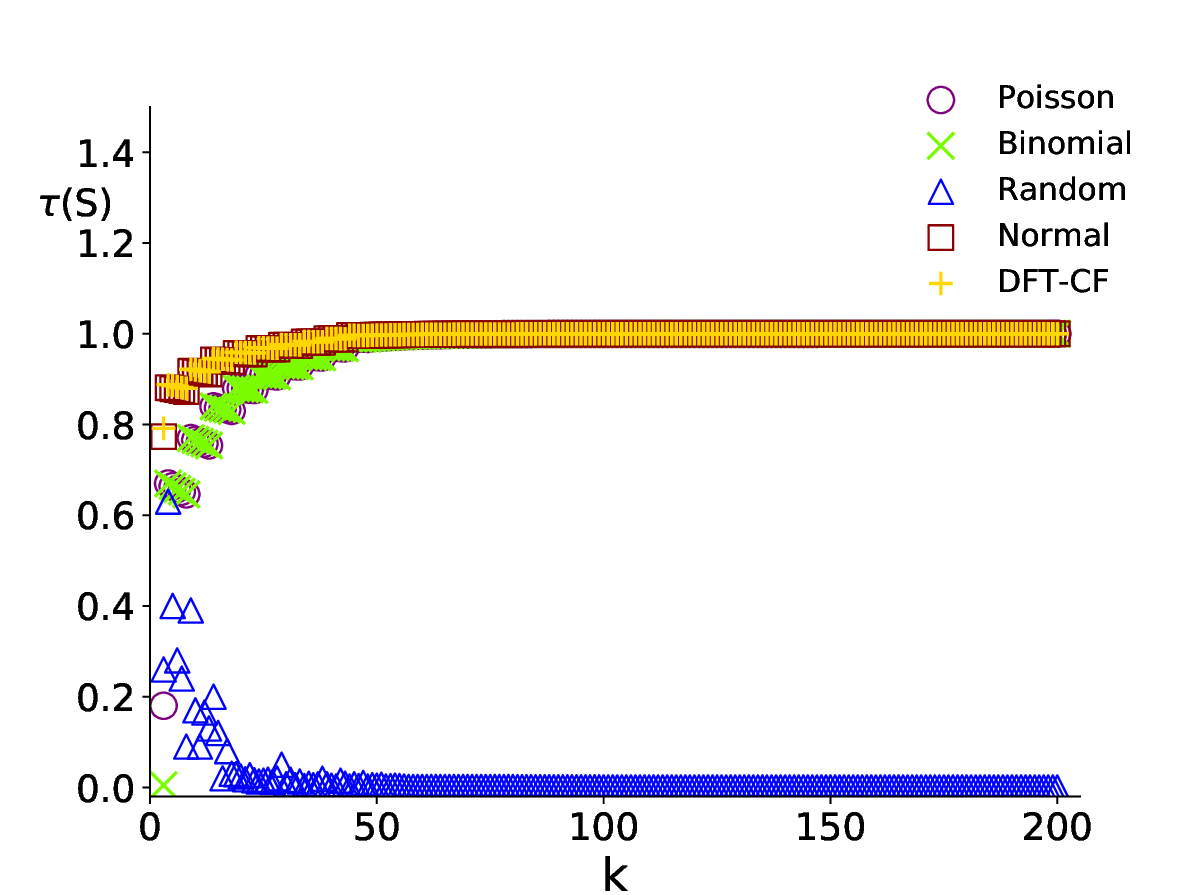}}
\caption{Testing on real data}
\label{fig:9}
\end{figure*}
In this subsection, we evaluated the proposed methods on real data sets from Foursquare. In particular, we select 10000 active workers (i.e. Foursquare users) from all the data collected. We evaluate sentiment of all the historical comments for each worker, and use average opinion sentiment value for this experiment. With this large data set, we examine the performance of the proposed algorithms with different settings of $\theta_0$, $\theta_1$ and $k$. In Fig.\ref{fig:9}, we use x-axis to denote the value of $k$, whereas $\theta_0$ and $\theta_1$ are set to be different portions of $k$. The baseline of these figures are the `Random' point set.\\
\indent First, we can observe that the proposed approximation-based methods significantly outperforms the random baseline. In particular, the advantage of proposals is evident when $\theta_0$ and $\theta_1$ are far from the $k/2$, such as Fig.\ref{fig:9}(a), \ref{fig:9}(d) and \ref{fig:9}(h). Comparatively, when they are close to $k/2$, the performance of random baseline becomes better, but still worse than our proposals. This phenomenon can be explained by the Central Limit Theorem \cite{Ref29} - the sum of 0-1 random variables (i.e. a Poisson Binomial distribution) is approximately a normal distribution, and the random baseline is more likely to pick the workers with probability close to the mean. So when the user’s demand is also close to the mean, the random baseline would have a better performance. When the user’s demand is far to the mean, randomly selecting workers is very unlikely to satisfy the user’s demand.\\
\indent Overall speaking, our proposal demonstrates very stable and outstanding performance. Moreover, when $k$ is fairly large, the user’s demand can be almost 100\% guaranteed. When $k$ is small, both the normal approximation and the exact DFT-CF method perform better than the Poisson approximation and the Binomial approximation. There are two main reasons for this phenomenon: (1) Both the Poission and Binomial approximation with the Backtracking Algorithms approximate $\tau(S)$ twice while the normal approximation with the simulated annealing algorithm approximates $\tau(S)$ once only, and there is no approximation for the exact DFT-CF method with the simulated annealing algorithm; (2) the order of worker selection affects the results of the Backtracking Algorithm, since in the Algorithm 2, the condition $\sum_{i=1}^kPr(t_i=1)>\Omega$ in line 4 determines whether the currently selected worker combination can continue to recursively from a qualified $S$, and it will be affected by the order of the data. Specifically, for different sorting of the same set of workers, when the first $k$ items are large enough, $\sum_{i=1}^kPr(t_i=1)>\Omega$ will be more likely to be judged as True, causing the currently worker combination and all $S$ combinations containing it to be removed. This will reduce the number of $S$ combinations traversed by the algorithm, thereby decreasing the possibility of searching for the optimal solution. And if the sum of the first $k$ items is too small, the number of $S$ combinations traversed by the algorithm will be the same as enumeration, which will affect the overall efficiency of the algorithm. However, the order of workers does not affect the result of the simulated annealing algorithm since the algorithm selects the workers randomly.
\subsection{Case Study}
\label{Case study}
\begin{figure}[ht]
\centering
  \includegraphics[scale=0.4]{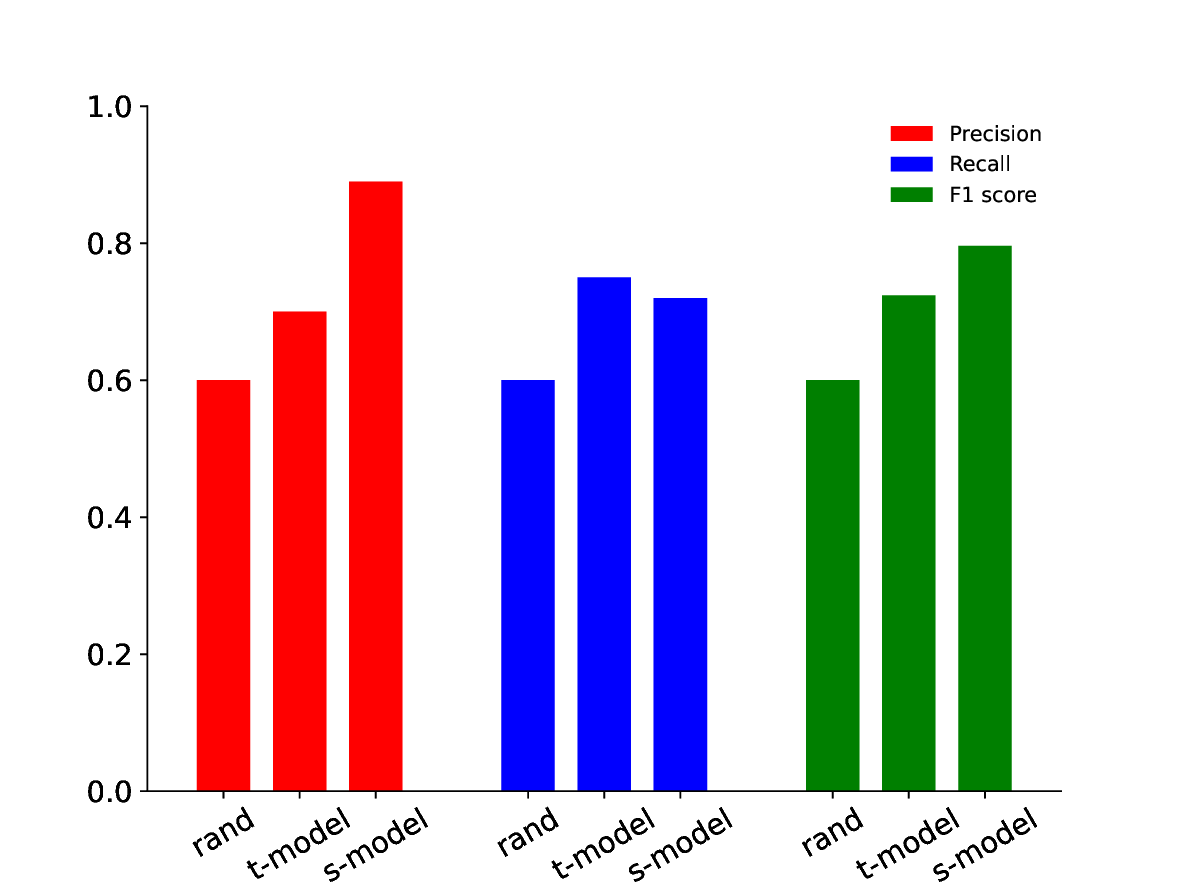}
\caption{Precision, Recall and F1 score on Case Study over AMT}
\label{fig:10}       
\end{figure}

We conducted a case study to exhibit the \textit{goodness} of crowds selected by our proposed models. In particular, we ask the crowds to produce pairwise comparisons for a number of restaurants. One thing worth noting is that the goodness of a crowdsourced result for restaurants is not \textit{absolute}. Nevertheless, in order to present a fairly objective evaluation, we carefully select 40 pairs of restaurants, such that each of them is consistently ranked by three different third-party systems, namely \textit{Yelp!} (http://www.yelp.com/ ), \textit{Michelin} (http://www.michelin.com/), as well as \textit{OpenRice} (http://www.openrice.com/). The pairwise comparisons agreed by all the systems are assumed to be the ground truth. \\ 
\indent We publish questions on Amazon Mechanical Turk (AMT), which is a widely used crowdsourcing marketplace. Each question contains two restaurants, and requires a worker to provide comments (at least 200 words) on each restaurant and decide which one is better. We accept 100 workers for each question. We apply the S-model and T-model on the data obtained from AMT, and select a subset of workers out of the 100 for each pair of restaurants. Specifically, we adopt the distance function detailed in section \ref{Pairwise experience-based diversity} for S-model; and use the sentiment analysis tool from Natural Language Toolkit (NLTK \cite{Ref5}) for the T-model. To aggregate crowdsourced answers, we use the majority as the crowd’s result. Moreover, for comparison, we randomly select the same number of workers, denoted by \textit{rand}.\\
\indent The size of the selected subset of workers is set to 11, 21,..., 51, and the proposed models consistently outperform \textit{rand}. Due to the page limit, we demonstrate the precision, recall and F1 score when the size is 21. In Fig.\ref{fig:10}, we use \textit{rand}, \textit{t-model} and \textit{s-model} to denote the results for random selection, T-model and S-model, respectively. From the experimental results, we can see that the proposed models achieve fairly high precision, recall and F1 score (70\%+). Besides, we observe that \textit{rand} has quite low precision, recall and F1 score, which indicate that the diversity of opinion is very important for constructing a crowd.
\section{RELATED WORK}
\label{Related Work}
\subsection{Crowd-based Queries}
\label{research related to crowdsourcing}
The recent development of crowdsourcing brings us a new opportunity to engage human intelligence into the process of answering queries (see \cite{Ref13} as a survey). Crowdsourcing provides a new problem-solving paradigm \cite{Ref8,Ref21}, which has been blended into several research communities. In particular, crowdsourcing-based data management techniques have attracted many attentions in the database and data mining communities recently. In the practical viewpoint, \cite{Ref15} proposed and develop a query processing system using microtask-based crowdsourcing to answer queries. Moreover, in \cite{Ref26}, a declarative query model is proposed to cooperate with standard relational database operators. In addition, in the viewpoint of theoretical study, many fundamental queries have been extensively studied, including filtering \cite{Ref25}, max \cite{Ref17}, sorting \cite{Ref22}, join \cite{Ref22,Ref23}, etc. Recently, some researchers considered new models with lower cost and higher efficiency than traditional models. For example, \cite{Ref44} proposed novel frameworks called SPR and SPR+ that are used to address the crowdsourced top-k queries to minimize the total monetary cost and \cite{Ref40} proposed a crowd-powered database system called `CDB' that tried to achieve the three optimization goals that are smaller monetary cost, lower latency and higher quality. Besides, crowdsourcing-based solutions of many complex algorithms are developed, such as categorization based on graph search \cite{Ref27}, clustering \cite{Ref16}, entity resolution \cite{Ref32,Ref34} and \cite{Ref42}, analysis over social media \cite{Ref10}, and tagging in social networks \cite{Ref12}, search engines \cite{Ref46}, trip planning \cite{Ref18}, pattern mining \cite{Ref6}, data management \cite{Ref45}, services like decision-making services \cite{Ref43} and service-oriented business \cite{Ref41}, etc. 
\subsection{Team Formation}
\label{research related to team formation}
Another related problem in the field of data mining is \textit{Team Formation Problem} \cite{Ref20}. Before taking diversity into consideration, previous Team Formation problems focus on satisfying the specific requirements of given tasks for certain skills which are possessed by different candidates experts. Normally, the cost of choosing one expert is also defined, e.g. influence on personal relationship and communication cost etc. Aside from using explicit graph constraints, some attempts of solving team formation problem are based on communication activities \cite{Ref9,Ref14}. In recent years, more and more methods and solutions have been proposed. For example, \cite{Ref51}, \cite{Ref49}, \cite{Ref50}, \cite{Ref47} respectively proposed different models or algorithms from the perspectives of member collaboration, communication costs among experts, integrating worker agency and team leaders. \cite{Ref48} also proposed a Decision Support System to assist multiple team formation in the context of software development that planned to be used in to the industry.\\ 
\indent The difference between Team Formation problem and ours is twofold. First, Team Formation mainly considers on individual capabilities, while we consider the crowd as a whole - the most capable workers may not make a wise crowd \cite{Ref31}. Second, we focus on the diversity of opinions of the crowd, which has not been addressed in the Team Formation problem. 
\subsection{Diversity of Opinions in Social Science}
\label{research related to diversity}
The importance of diversity of opinions for crowdsourcing is already well studied in the field of social science. In particular, \cite{Ref23} is known as one of the most representative book in the field. It highlights the importance of cognitive diversity for collective problemsolving (where diversity trumps ability), and takes a complex subject, moves beyond metaphor and mysticism and politics and places the claims of diversity’s benefits on a solid intellectual foundation.\\
\indent To our best knowledge, this is the first work of algorithmic study on a how to construct a wise crowd with the consideration of the diversity of opinion.
\section{CONCLUSION AND FUTURE WORK}
\label{Conclusion and Future work}
In this paper, we study how to construct a wise crowd with the consideration of diversity of opinions. In particular, two basic paradigms for worker selection is addressed - building a crowd waiting for tasks to come and selecting workers for a given task. Accordingly, we propose Similarity-driven (S-model) and Task-driven Model (T-model) for these two paradigms. Under both of the models, we propose efficient and effective algorithms to enlist workers with a budgeted constraint. We have verified the solutions with extensive experiments on both synthetic datasets and real data sets. The experimental studies demonstrate that the proposals are robust for varying parameters, and significantly outperform the baselines.\\
\indent It is not suitable to combine S-model and T-model as they are devised to estimate the diversity of crowd under two very distinct scenarios. The S-model is established to form a diverse combination of workers that can handle as many tasks as possible, and the scope of measuring the similarity of the S-model generally involves a wide range of tasks involved in the T-model, so the similarity in the S-model cannot be directly used in the T-model. For example, consider two workers whose similarity is 0.9, and the probability that they have a positive attitude towards a certain task is 0.9. The S-model still has a high probability of not selecting them, but when the positive demand is high, the T-model has a very high probability of selecting them, which leads to different results between the two models. \\
\indent Our methods have the potential to be combined with other existing methods in two directions. The first is to combine existing crowdsourcing methods with ours, which can be seen as a secondary screening of candidates to construct a stronger crowdsourcing ensemble. For example, algorithms mentioned in \cite{Ref57} can be used to pre-sort candidates relevant to tasks. Using this algorithm, skill scores of candidates can be computed, filtering a batch of candidates suitable for specific tasks, and then using the S-model for a diversified selection of opinions as much as possible. Alternatively, this algorithm can be placed in the next step of the S-model, computing among the candidates selected from the S-model those relevant to specific jobs. The second direction is to use other methods that provides the input data, including similarity in the S-model and the probabilities in the T-model. For instance, \cite{Ref58} proposed a model that infers users' sentiments and opinions based on long-term historical information. We can utilize this model by analyzing the historical text information related to tasks that workers have previously posted (such as social media updates) to predict the approximate attitudes (or probabilities) held by the workers. We can then use this information as a reference to calculate the probability of workers holding a positive attitude in the T-model and proceed with the selection process accordingly.
\\
\indent There are many further research directions to explore. One immediate future direction is how to consider the different influence of workers for the diversity of opinions. The influence may diminish the range of opinions, and polarize people’s opinions making group feedback less reliable in guiding decision-makers. Influencers tend to improve people's confidence, but this so-called `confidence effect' will boost an individual's confidence, while at the same time, decrease their accuracy. Another interesting dimension is to differentiate the cost for recruiting different workers, then the problem is to minimize the total cost while fulfilling the requirement of diversity. Besides, we are interested in designing better similarity/distance functions for our T-model.

%
%




\section{APPENDIX}
\label{Appendix}
\subsection{POISSON AND BINOMIAL APPROXIMATION}
\label{T-model approximations}
In Section 3, we use Poisson distribution and Binomial distribution to approximate Poisson Binomial distribution. Here, we conclude the quality of approximation in \cite{Ref30,Ref29}.\\
\indent Let $X_1,X_2,...,X_n$ be a set of Bernoulli trials such that $Pr(X_j = 1) = p_j$ and $X = \sum_{j=1}^nX_j$. Then $X$ follows a Poisson Binomial distribution. Suppose $\mu = E[X] = \sum_{j=1}^n p_j$. The probability of $X = i$ and $X \leq i$ can be approximated by the probability density function (PDF) and cumulative mass function (CMF) of Poisson distribution and Binomial distribution.\\
\indent \textbf{Poisson Approximation:}
\begin{equation}
    Pr(X \leq i) \approx F_P(i,u) = \frac{\Gamma(i+1,u)}{i!}e^{-u}
\end{equation}
\cite{Ref30} provides an upper bound of the error of the approximation:
\begin{equation}
    |Pr(X \leq i) - F_p(i,u)| \leq min(\mu-1 \wedge 1)\sum_{j=1}^{n}p_j^2
\end{equation}
for $i = 0,1,2,...,n$. Clearly, this upper bound of the error is greater than or equal to 0. When $\mu \in [0,1]$
\begin{equation}
    |Pr(X \leq i) - F_p(i,u)| = \sum_{j=1}^np_j^2 \leq \sum_{j=1}^np_j \leq 1
\end{equation}
When $\mu \in [1,+\infty)$
\begin{equation}
    |Pr(X \leq i) - F_p(i,u)| = \frac{\sum_{j=1}^np_j^2}{\sum_{j=1}^np_j} \leq \frac{\sum_{j=1}^np_j}{\sum_{j=1}^np_j} = 1
\end{equation}
So, in ether case:
\begin{equation}
    0 \leq |Pr(X \leq i) - F_p(i,u)| \leq 1
\end{equation}
\indent \textbf{Binomial Approximation:}\\
\indent In \cite{Ref29}, the metric of error is defined as
\begin{equation}
    d_{err} =\frac{1}{2}\sum_{i\in\mathbb{Z}}|f_B(X=i)-B_i(i;n,p)|
\end{equation}
By using Binomial distribution $B_i(i;n,p)$ approximate the distribution of $X$, where $p = \mu/n$, we have
\begin{equation}
    d_{err} \leq \frac{1-p^{n+1}-(1-p)^{n+1}}{(n+1)p(1-p)}\sum_{i=1}^n(p_i-p)^2
\end{equation}
\end{document}